\documentclass[12pt,tightenlines,showpacs,preprintnumbers,superscriptaddress,nofootinbib]{article}
\pdfoutput=1 
\usepackage{amsmath,amssymb,amsthm}
\usepackage{bbm,latexsym}
\usepackage{graphicx}
\usepackage{rotating} 
\usepackage{amsfonts} 
\usepackage{xcolor}
\usepackage[numbers,sort&compress]{natbib}
\usepackage{authblk}
\usepackage{hyperref}

\newcommand{\be}{\begin{equation}}
\newcommand{\ee}{\end{equation}}
\newcommand{\ba}{\begin{array}}
\newcommand{\ea}{\end{array}}
\newcommand{\bea}{\begin{eqnarray}}
\newcommand{\eea}{\end{eqnarray}}
\newcommand{\balg}{\begin{align}}
\newcommand{\ealg}{\end{align}}
\newcommand{\bit}{\begin{itemize}}
\newcommand{\eit}{\end{itemize}}
\newcommand{\trm}[1]{\textrm{#1}}

\newcommand{\Mpc}{\trm{\Mpc}}
\newcommand{\yr}{\trm{\yr}}
\newcommand{\eV}{\trm{\eV}}

\newcommand{\Fi}{{16}_{Fi}}
\newcommand{\Fj}{{16}_{Fj}}

\begin{document}

\title{Neutrino mixing and masses in SO(10) GUTs with  hidden sector and flavor symmetries}

\setcounter{footnote}{2}
\author[1]{Xiaoyong Chu\thanks{xchu@ictp.it}}
\author[2,1]{Alexei Yu.\ Smirnov\thanks{smirnov@mpi-hd.mpg.de}}
\affil[1]{{\small \it International Centre for Theoretical Physics, Strada Costiera 11, I-34100 Trieste, Italy}}
\affil[2]{{\small \it Max-Planck-Institute for Nuclear Physics, Saupfercheckweg 1, D-69117 Heidelberg, Germany}}

\date{}

\maketitle

\begin{abstract}

We consider the neutrino  masses and mixing  in the framework
of SO(10) GUTs with hidden sector consisting of fermionic and bosonic
SO(10) singlets  and flavor symmetries.
The framework allows to disentangle the {\it CKM physics}
responsible for the CKM mixing and different mass hierarchies of
quarks and leptons and the {\it neutrino new physics} which produces
smallness of neutrino masses and large lepton mixing.
The framework leads naturally to the relation 
 $U_{PMNS} \sim V_{CKM}^{\dagger} U_0$,  where structure of $U_0$
is determined by the flavor symmetry.
The key feature of the framework is that apart from the Dirac
mass matrices $m_D$, 
the portal mass matrix $M_D$ and the mass matrix of singlets $M_S$
are also involved in generation of the lepton mixing. 
This opens up new possibilities to realize the flavor symmetries  and explain the data.
Using $A_4 \times Z_4$  as the flavor group, we systematically explore
the flavor structures  which can be obtained in this framework
depending on field content and symmetry assignments.
We formulate additional conditions which lead to $U_0 \sim U_{TBM}$ or $U_{BM}$.
They include (i) equality (in general, proportionality) of the
singlet flavons couplings,
(ii) equality of their VEVs;
(iii) correlation between VEVs of singlets and triplet,
(iv) certain VEV alignment of flavon triplet(s).
These features can follow from additional symmetries or be remnants of further unification.
Phenomenologically viable schemes with minimal flavon content and minimal number of couplings are
constructed.

\vspace{.4cm}

\end{abstract}

\maketitle

\vspace{.4cm}

\section{Introduction} \label{chapt:Intro}

There are several 
hints in favor of a framework with Grand unification, hidden sector at the string/Planck scale, and flavor symmetries acting both in visible and hidden sectors:

1. Grand Unification Theories (GUT) are still one of the highly motivated 
and  very appealing scenarios of physics beyond the Standard Model (SM)~\cite{Pati:1973uk,Georgi:1974sy}. 
In particular, the GUT models based on SO(10)~\cite{Georgi:1974my,Fritzsch:1974nn} 
beside the gauge coupling unification and charge quantization 
accommodate exactly  the quantum numbers of all known SM fermions, 
plus right-handed (RH) neutrinos.   
This can not be accidental.  
The SO(10) GUT also provides natural realizations of the high scale seesaw 
mechanism~\cite{Minkowski:1977sc,Yanagida:1979as,GellMann:1980vs,Glashow:1979nm,Mohapatra:1980yp}, 
 and therefore explanation of the smallness of neutrino masses. 

2. In fact, the required mass scale of the RH neutrinos from the seesaw mechanism, 
$M_N \sim m_D^2/ m_\nu \sim (10^8 - 10^{14})$ GeV, where $m_D$ is a typical Dirac mass of the quarks 
and charged leptons,  is somehow below the GUT scale.  Nevertheless, the scale of $M_N$  
can be made of the GUT scale $M_{GUT}$ 
via the seesaw mechanism, if additional fermion singlets $S$ exist with masses at the 
Planck scale $M_{Pl}$: $M_N \sim M_{GUT}^2/ M_{Pl}$. Consequently, 
the double seesaw mechanism  is realized for light neutrino masses~\cite{Mohapatra:1986bd}.

3. Generically,  the unification is at odd with strong difference of mixings 
and masses of quarks and leptons.  
Partially, the difference of mixings can be related to the smallness 
of neutrino mass~\cite{Smirnov:1993af,Smirnov:1995hk,Tanimoto:1995nq} 
and the latter can originate again from the seesaw 
mechanism. 

4. 
The approximate tri-bimaximal (TBM) lepton mixing~\cite{Harrison:2002er}
can be considered as an indication of existence of the flavor symmetry
behind~\cite{Harrison:2002kp,Harrison:2003aw,Harrison:2004uh}. 
Furthermore, simple discrete groups, like $A_4$~\cite{Wyler:1979fe, Branco:1980ax}, have certain properties which 
correspond to  features of the TBM mixing~\cite{Ma:2001dn, Babu:2002dz, Hirsch:2003dr, Altarelli:2005yp, King:2006np, Koide:2007kw}.  
In SO(10),  realization of such a symmetry faces serious
problems in view of strongly different  mixing pattern 
in the quark sector~\cite{Barr:2001vj,Ma:2002yp}. For  some realizations with $A_4$ in SO(10) GUT, 
see e.g. \cite{Ma:2006wm, Morisi:2007ft, Bazzocchi:2008rz}.

5. Although being strongly different, the quark and lepton mixings can be related.  Indeed, 
experimental data 
confirm the following connection between the quark,  $V_{CKM}$, and lepton, $U_{PMNS}$,   
mixing matrices~\cite{Giunti:2002ye,Minakata:2004xt,Cheung:2005gq,Xing:2005ur,Datta:2005ci,
Antusch:2005ca,Antusch:2005kw,Harada:2005km,Farzan:2006vj}  
\be
U_{PMNS} \simeq V_{CKM}^\dagger U_0.     
\label{ql:comp}
\ee
Here $U_0$ is the matrix of special form, like the TBM mixing matrix $U_{TBM}$ 
or the bi-maximal (BM) mixing $U_{BM}$~\cite{Vissani:1997pa, Fukugita:1998vn, 
Barger:1998ta, Davidson:1998bi, Mohapatra:1998ka}, dictated by symmetry 
and related to some new structures of theory which affect the lepton sector only. 
Eq.~\eqref{ql:comp} gives a relation between 
the 1-3 mixing, 2-3 mixing and CKM mixing, which is in a good agreement with data \cite{Ludl:2015tha}. 
In the lowest-order (only Cabibbo angle $\theta_C$ in $V_{CKM}$, and maximal 2-3 mixing) the relation 
reduces to 
$\sin^2 \theta_{13} = 0.5 \sin^2 \theta_{C}$ \cite{Minakata:2004xt,King:2012vj}. 

The relation (\ref{ql:comp}) implies Grand unification, which transfers information about mixing 
from the quark to lepton sector.

6. The double seesaw mechanism allows to explain  difference of mixing patterns  
in the quark and lepton sectors and reproduce the relation (\ref{ql:comp}).  
It  reveals new possibilities to realize  flavor symmetries. 
Striking differences in the quark and lepton sectors follow essentially from 
the mixing of new fermion singlets $S$ with neutrino (left- and right-handed) 
components of $16$-plet,  thus providing the main source of  the lepton mixing.

Origin of singlets $S$ can be related to further extension of the gauge symmetry:
SO(10) $\rightarrow E_6$, in which new singlets are additional neutral components of
$27$-plets~\cite{Gursey:1975ki}. Alternatively,  singlet fermions $S$ can be components 
of the hidden sector  composed of the fermionic and bosonic singlets of SO(10).
Hidden sector may have its own gauge symmetries. 
Sterile neutrinos,  if exist, particles of dark matter, 
axions, {\it etc.}, can also originate from this sector.

It seems that the existing data on masses and mixing 
are too complicated to be explained all at once, from a few principles and  a 
few parameters. 
So, according to Eq.~(\ref{ql:comp}), 
probably two different (but somehow connected) types of new physics are involved:
the ``CKM new physics'' responsible for the CKM mixing
and difference of  hierarchies of Dirac masses,  and the ``neutrino new physics",
which explains smallness of neutrino masses and large mixing
with its form dictated by  symmetry. 
The neutrino new physics is related to the hidden sector and
properties of the portal interactions which
connect the visible sector and the hidden one~\cite{Lee:1956qn,Salam:1957st,Kobzarev:1966qya}.

This framework  has been explored to some extent  previously.
That includes (i) the seesaw enhancement of lepton mixing due to 
coupling with new singlets \cite{Smirnov:1993af}, 
(ii) screening of the Dirac structures, 
which means that the mixing of light neutrinos is essentially related to 
mixing in the singlet sector $S$: $m_\nu  \propto M_S$  
\cite{Smirnov:1993af, Lindner:2005pk},  
(iii)  stability of this screening 
with respect to renormalization group (RG) running~\cite{Lindner:2005pk}, 
renormalization of the relation (\ref{ql:comp}) \cite{Schmidt:2006rb}, 
(iv) symmetry origins (groups $T_7$ and $\Sigma_{81}$~ have been used) 
of the screening, as well as  partial screening \cite{Hagedorn:2008bc},  
(v) lepton mixing from the symmetries of the hidden sector~\cite{Ludl:2015tha}.

In this paper we  further elaborate on the framework with 
emphasis on new mechanisms of getting 
flavor structures from symmetries. 
We systematically explore the fermion masses and mixing matrices which can be obtained in the framework. 
Interplay of the flavor and GUT symmetries is crucial. 
We mainly discuss here the symmetry aspects and formulate 
conditions of the vacuum alignment which eventually 
should be realized in complete model buildings.

The paper is organized as follows. In sect.~\ref{chapt:frame} we study 
properties of the  mass matrices involved in generation of the neutrino 
masses and mixing in this framework. 
We elaborate on the main elements  and building blocks of the schemes of lepton mixing. 
In sect.~\ref{sect:trip} we present the simplest viable  schemes 
with fermion singlets $S$  transforming under three-dimensional representation {\bf 3} of $A_4$. 
In sect.~\ref{sect:sing} the schemes are considered  with $S$  transforming under one-dimensional 
representations ${\bf 1, 1', 1''}$ of $A_4$. Sect.~\ref{sect:general} is devoted to 
various generalizations: effect of linear seesaw, contribution from additional singlets, 
elements of the CKM physics. We summarize our results in sect.~\ref{sect:conclusion}.  
Some technical details are presented  in the Appendix. 

\section{Neutrino mass and mixing matrices} \label{chapt:frame}

\subsection{The framework}

Let us first summarize main ingredients of the framework  outlined in the introduction.  
We consider the SO(10) gauge model with three generations 
of fermions  and  the following additional features:     

\begin{itemize}

\item
All the chiral fermions (including the  right-handed neutrino)  of each generation 
are accommodated in a 16-dimensional spinor representation, $\Fi$ ($i=1,2,3$).

\item  
Three or more singlet fermions $S_i$ exist.  They mix with neutrinos due to 
Yukawa interactions with $16$-plets of Higgs scalars $\overline{16}_H$. 

\item 
Only low-dimension representations of Higgs fields, $10_H$, $\overline{16}_H$ and $1_H$,  
give masses to  fermions. High dimensional representations $126_H$ and $120_H$ are absent. 

\item 
Approximate flavor symmetry exists with fermions $\Fi$, $S_j$   carrying 
non-zero flavor charges.  For definiteness we consider the simplest 
discrete group which has irreducible 
three-dimensional representations, $A_4$. Various features obtained from this  symmetry 
can be also reproduced with other groups.

\item
Three generations of the matter fields, $\Fi$,  
transform as triplet of  $A_4$ in most of the cases.  
We explore different symmetry assignments for three singlets $S_i$, 
$({\bf 3}, {\bf 1}, {\bf 1'}, {\bf 1''})$.  

\item 
A number of flavons (scalar fields, SO(10) singlets) exist,  
which have non-trivial 
transformations with respect to $A_4$ and acquire 
vacuum expectation values\,(VEV), thus breaking the symmetry.  

\item 
The Higgs multiplets, $10_H$ and ${16}_H$, are singlets of 
$A_4$, and therefore factor out of the flavor structure in the simplest cases. 
This reduces the complexity 
in obtaining certain form of mass matrices and
mixing from flavor symmetries. 
Still, $10_H$ and ${16}_H$  may carry charges of additional symmetries, 
in which case they will affect the flavor.

\end{itemize}

We do not specify here the SO(10) symmetry breaking, 
although the corresponding Higgs multiplets can also take part 
in generation of the fermion masses. 
Rather than constructing complete  models,
we will explore systematically the  flavor
structures which can be obtained in the formulated framework.

\subsection{Yukawa couplings and mass matrices} \label{sect:IIB}

The Lagrangian  has three types of Yukawa couplings,  
\be
{\mathcal L} = {\mathcal L_{D}}+{\mathcal L_{NS}}+{\mathcal L_{S}},
\label{eq:larg}
\ee
which produce masses of fermions:

1). The {\it visible} (Dirac) sector  Yukawa interactions 
\be
{\mathcal L}_{D}\, =  Y_{10}^{ij a} \cdot \Fi \Fj \, 10_H^a \,   
\label{Lagr:general}
\ee
generate after symmetry breaking the usual Dirac masses of fermions. 
With just one  $10_H$  
these couplings produce  Dirac mass matrices of the up-type quark, down-type quark, 
charge lepton mass matrices and neutrinos   
of the same form:  
\be
m_u = m_{D} =  Y_{10} v_u, ~~~
 m_d = m_l  =  Y_{10} v_d,
\label{eq:m-dir}
\ee
where $v_u$ $v_d$ are the VEVs of $10_H$.  
No quark mixing is generated at this level. Nevertheless, such a mixing can be produced if 
two (or more) different  $10_H^a$ $(a = u, d)$  are introduced  
(see sect.~\ref{chapt:CKM}).

2). The {\it portal} interactions, 
\be 
{\mathcal L}_{NS} = Y_{16}^{ij}\cdot \Fi S_j \, \overline{16}_H, 
\label{portal-int}
\ee 
generate mass matrices of Dirac type for the LH and RH neutrinos from $\Fi$ 
and  singlets $S_j$ 
$$
m' = Y_{16} v_L,~~~ M_D = Y_{16} v_R. ~~~ 
$$
Here $v_L$, $v_R$  are  the VEVs of the $SU(2)_L$ and $SU(2)_R$ doublets from 
$\overline{16}_H$. 
As long as  only one Higgs 16-plet  generates 
$M_D$ and $m'$, the two mass matrices are proportional to each other: 
\be
m' = \frac{v_L}{v_R} M_D. 
\label{eq:propto}
\ee  

3). The {\it hidden sector} interactions, 
\be 
{\mathcal L}_{S} = Y_{1}^{ij}\cdot S_i S_j \, 1_H + M^{0\,ij}_S S_i S_j,
\label{hidden-int}
\ee 
produce the mass matrix of singlets  
$$
M_S = Y_1 \langle 1_H \rangle  + M_S^0. 
$$
Here $1_H$ fields are singlets under both flavon symmetry and $SO(10)$, which may not be necessary and are assumed to be absorbed into flavons from now on. 
The masses of singlets can be  generated 
at the renormalizable level,  and also bare mass terms $M_{S}^0$  can appear 
for some or all singlets if it is not forbidden by flavor symmetry.

\vspace{.2cm}

Notice that here the Yukawa couplings are effective couplings 
-- functions of VEVs of flavon fields, $\Phi$:  
$$
Y_{k}^{ija} = Y_{k}^{ija}(\langle \Phi \rangle). 
$$ 
If $\langle \Phi \rangle \ll \Lambda_f$,  where 
$\Lambda_f$ is a scale of flavor physics, 
the constants  can be expanded as  
$$
Y_{k}^{ija} = y_{k}^{ija} + \left[y_k {\langle \Phi \rangle \over \Lambda_f}\right]^{ija} + 
\left[y_k^2 {\langle \Phi \rangle^2 \over \Lambda_f^2}\right]^{ija}  + ..., ~~~~k = 10,~16,~1.  
$$    
In what follows we will take the lowest-order approximation in 
$\langle \Phi \rangle /\Lambda_f$.  

The mass matrix of neutral leptons generated by interactions   
(\ref{Lagr:general}),   (\ref{portal-int}) and   (\ref{hidden-int})
in the basis $(\nu, N, S)$, 
where  $N \equiv \nu^c$,   equals
\begin{equation}
\begin{pmatrix}
0        &m_D          & m' \\
m_D^T    &0            & M_D\\
m'^T    &M_D^T     & M_S\\
\end{pmatrix}\,.      
\label{General:Mnu}
\end{equation}
Given the Yukawa interactions introduced above, 
the matrices $m_D$ and $M_S$ are symmetric:  $m_D = m_D^T$ and 
$M_S = M_S^T$, while $m'$ and $M_D$  may have a non-symmetric form. 
The mass terms are defined as $\nu^T m_D C^{-1} N$, $N^T M_D C^{-1}S$, {\it etc.}, and 
$C \equiv i \gamma^0 \gamma^2$. 
For 
$$
m_D, ~ m' \ll M_D, ~~ M_S 
$$
the diagonalization of  (\ref{General:Mnu}) leads to the light neutrino mass matrix
\begin{equation}
m_\nu \simeq  \,  m_D \, {1\over M_D^T} \, {M_S} \, {1\over M_D} m_D^T \,
- \left(m_D {1\over M_D^T} m'^T +  
m' {1\over M_D}  m_D^T\right), 
\label{neutrino:2mass}    
\end{equation}
where the first term is the double seesaw  contribution, 
and two others correspond to the linear seesaw contribution.  
In the case of a single $\overline{16}_H$, according to Eq.~\eqref{eq:propto} 
the mass matrix (\ref{neutrino:2mass}) becomes  
\begin{equation}
m_\nu \simeq  \,  m_D \, {1\over M_D^T} \, 
{M_S} \, {1\over M_D} m_D^T \, - {2 v_L \over v_R } m_D.    
\label{numass3}
\end{equation}
Given that $m_\nu \sim$ O(0.1)\,eV and $v_L < \text{246\,GeV}$,
it is possible to define a critical mass scale  
\begin{equation}
M_{cr} = 4 \times 10^{14}  \,\left({v_L \cdot Y_{16}\over \text{246\,GeV}}
\right)^2 ~\text{GeV}\,,
\label{eq:crit}
\end{equation}
so that for $M_S \ll M_{cr}$ 
the "linear-seesaw"  dominates, whereas for  $M_S \gg M_{cr}$, the "double-seesaw" 
gives the main contribution. 
In our framework with $M_S \sim M_{Pl}$  
the "double-seesaw"  dominates and  
the "linear-seesaw" term may lead to small corrections.

\vspace{.2cm}

According to Eq.~(\ref{eq:m-dir}) 
and  $m_D \propto  m_l$,   
no mixing is produced by the Dirac mass  matrices.  
In contrast,  due to extended structure of the mass matrix 
of neutral leptons Eq.~\eqref{General:Mnu},  $m_\nu$ 
differs from $m_l$,  and so the lepton mixing is generated. 
In the basis where the Dirac mass matrices are diagonal, $m_D = m_D^{diag}$,  
the origin of lepton mixing is the structure 
\be
m_D^{diag} M_D^{-1\, T} M_S  M_D^{-1} m_D^{diag}.  
\label{double:seesaw}
\ee

In the lowest order, the Majorana mass term of right-handed neutrinos, $N^T C^{-1} N $, is absent, 
since we do not introduce $126$ Higgs multiplet.  
Such a term can  be generated by higher-dimension operator   
 $ (1/\Lambda)  \Fi \Fj (16_H 16_H)^{ij}$,  
or effectively generated by the decoupling of heavy particles with mass 
$\sim \Lambda$.  
It then gives $M_N = v_R^2/\Lambda$.

In our framework,   the mass term $N^T C^{-1} N $ appears after integrating out 
$S$, which gives
\be
M_N =   M_D^T  {M_S}^{-1}  M_D =   v_R^2  Y_{16}^T {M_S}^{-1} Y_{16}.
\label{eq:rhnmass}
\ee
We assume that this contribution to $M_N$ dominates. Other contributions  either
are forbidden  by certain auxiliary symmetry or have 
the same flavor structure as  Eq.~(\ref{eq:rhnmass}). 

Notice that since in our framework  there is $m_D \ll M_D$, 
the RG running of Yukawa couplings may modify 
the expressions for neutrino masses non-negligibly, 
and thus may need to be taken into 
account~\cite{Chankowski:2001mx, King:2000hk, King:2000ce, Antusch:2002rr, Stech:2003sb}.

\subsection{Screening and partial screening}

The double seesaw allows to realize the screening of the Dirac structures 
\cite{Smirnov:1993af, Lindner:2005pk}, 
which will be used in many examples presented in this paper.   
If 
\be 
m_D \propto M_D^T,  
\label{eq:scr}
\ee 
then according to Eq.~(\ref{numass3})
\be
m_\nu \propto M_S, 
\label{eq:screen}
\ee
{\it i.e.} structure of the light neutrino mass matrix is given 
by that of $M_S$, thus disentangling properties of  
neutrinos and quarks completely. 
The form  of $M_S$ can strongly differ from those of $m_D$ and   $M_D$,
thus leading to very different lepton mixing. 

Suppose that the  screening condition is fulfilled in the basis where 
$m_D \propto M_D^T$ are diagonal due to some certain symmetry, {\it i.e.} in the flavor basis, 
and that in this basis the singlet mass matrix is $M_S$. 
Then  according to (\ref{eq:screen}) the mixing is determined exclusively by $M_S$.

If the screening condition \eqref{eq:scr} is fulfilled in some other basis 
(again determined by symmetry) where 
$m_l = m_D$ is   non-diagonal and the singlet matrix is $M_S'$, 
then $m_\nu \propto M_S'$.  But now the lepton mixing  gets contributions 
both from $m_D$ and $M_S'$.  
We present schemes which realize both possibilities.

Let us consider the screening condition in more details. 
We define the unitary transformations $U_l$, $V_R$, 
$V_S$  and $V_N$ which diagonalize the mass matrices $m_D$ and $M_D$, 
so that 
$$
m_D = U_l m_D^{diag} V_R^{\dagger}, ~~~~ M_D^T = V_S M_D^{diag} V_N^\dagger.
$$
Then the  screening matrix (factor) equals  
$$
m_D (M_D^T)^{-1} = U_l m_D^{diag} V_R^{\dagger} V_N  (M_D^{diag})^{-1} V_S^\dagger.    
$$
The screening  (\ref{eq:scr})  implies that 
$$
V_R = V_N, ~~~~ m_D^{diag} \propto M_D^{diag},  ~~~~~U_l  =  V_S.   
$$
If only the first condition (the same rotations of the 
RH neutrino components) and the second one are  fulfilled 
we obtain 
\be 
m_D (M_D^T)^{-1} = U_l V_S^\dagger \neq I. 
\label{partialscr}
\ee
This partial screening gives additional contribution to the neutrino  mixing. 
Due to symmetry the matrices $U_l$  and  $V_S$ can be related in such a way 
that Eq.~\eqref{partialscr} produces a matrix of special form, 
which then leads to the required form of total mixing matrix (see sect.~\ref{TBM:33mixing}).    
The  screening can be due to further embedding of singlets $S$ in extended gauge symmetry 
or due to common flavor symmetries. 

Some additional physics (which we call the CKM physics)
can lead to the misalignment of mass matrices for
up- and down- type quarks. Due to GUT similar misalignment is expected between 
the neutrino Dirac matrix $m_D$ and the charged lepton matrix 
$m_l \neq m_D$. Then in the presence of screening 
$m_D \propto M_D^T = diag$ the PMNS mixing gets  
the CKM type contribution $\sim V_{CKM}$ from $m_l$ and another one,  
$U_S$,  from $M_S$:
$$
U_{PMNS} = V_{CKM}^{\dagger} U_S. 
$$
This allows to maximally disentangle the
CKM and neutrino new physics and realize the scenario 
in Eq.~\eqref{ql:comp}.

\subsection{Symmetry and flavor structures} \label{sect:IID}

The flavor symmetry  imposes restrictions 
on the Yukawa  couplings $Y_k^{ij}$, and thus  forbids some of them 
or leads to relations between them.  
For definiteness we use  $A_4$ flavor group (see Appendix.~\ref{a4group} for details). 
Various structures we obtain in this paper can be reproduced in models
with other symmetry groups which  include representations
${\bf 3}, \, {\bf 1}, \,  {\bf 1'}, \, {\bf 1''}$.
On the other hand, with additional symmetries it is possible to produce
in our framework features from other symmetry groups. 

We introduce flavon triplets $\xi$ and singlets 
$\vec{\varphi} = (\varphi,~ \varphi',~ \varphi'')$,  
which acquire the VEV $\langle \xi  \rangle \equiv (u_1, u_2, u_3)$ and   
$(\langle \varphi \rangle, \langle \varphi' \rangle,  \langle \varphi'' \rangle)  \equiv (v, v', v'')$. 
The general form of a fermion mass term
is $ F_i F_j$, where the fermionic multiplets 
$F_i = 16_F, \,\, S$. Depending  on symmetry assignment
of $F_i$  and for general sets of the flavon fields,  
which include the flavon triplets
$\xi$ and singlets $\vec{\varphi}$, 
we obtain three types of mass matrices:

1.  $ F_1 \sim {\bf 3}$,  $F_2 \sim {\bf 3}$. 
The most general coupling with flavons invariant under $A_4$ 
is 
\begin{equation}
\label{mass:form33}
m_0 (F_1 F_2) +  \vec{y} (F_1 F_2)_\varphi \vec{\varphi}   +  h (F_1 F_2) \xi,   
\end{equation}
where we use  notations 
\be
\vec{y}~  (F_1 F_2)_\varphi   \vec{\varphi}
\equiv y (F_1 F_2) \varphi +  y' (F_1 F_2)'' \varphi'  +  y'' (F_1 F_2)' \varphi''.  
\label{notation1}
\ee
Here $\vec{y} \equiv (y, y', y'')$ and $(Q)_\varphi$     
($Q =  F_1 F_2$ in this case)  is the field operator  which 
produces one-dimensional representations of $A_4$ together with  a given $\varphi$.

The mass term $m_0(F_1 F_2)$  without flavons  can be  
forbidden by symmetries. The interactions (\ref{mass:form33}) lead to the mass terms 
$F_1 M_{33} F_2^T + h.c.$  with  
\begin{equation}
M_{33} =  M_{33}^{diag} +  M_{33}^{off} =  m_0 {\mathbbm 1} +  M_{33}^{\varphi} + M_{33}^{\xi}, 
\label{diag-off}
\end{equation}
where the off-diagonal part is generated by flavon triplet,  
whereas the diagonal one  
$$
M_{33}^{diag} = m_0 {\mathbbm 1} +  M_{33}^{\varphi} 
$$
is  due to coupling with the flavon singlets of $A_4$ and flavonless contribution. 
Three terms in (\ref{diag-off}) are generated by different flavons.  
Depending on additional symmetry assignments  
one, or two, or all three terms can contribute. 
Explicitly, 
\begin{equation}
M_{33}^{\varphi} =
\begin{pmatrix}
y v + y' v' +y'' v''   & 0  &  0\\
0      &   y v + \omega y' v' + \omega^2 y'' v''    & 0 \\
0   & 0   &    y v + \omega^2  y' v' + \omega  y'' v''
\end{pmatrix}.   
\label{33diag}
\end{equation}
Notice that for all equal values ($y v =  y' v' =  y'' v''$),  
$M_{33}^{diag} = diag(m_0 + 3 y v, m_0, m_0)$, which can be a good 
first approximation in some cases.  For real parameters $\mu \equiv m_0 + y v$,  $\mu' \equiv  y' v'$ and $\mu''  \equiv y'' v''$
two eigenstates are degenerate: 
$$
|M_ 2|  = |M_3| = 
[ (\mu + \mu' + \mu'')^2 - 3(\mu \mu' + \mu' \mu'' + \mu \mu'')]^{1/2}. 
$$
This important feature together with certain forms of the off-diagonal part can lead to maximal mixing. 
General situation is analyzed in the Appendix.~\ref{app:eigenvalues}.

In the absence of the bare mass $m_0$, the eigenvalues of matrix 
(\ref{33diag}) can be written as 
\be 
(M_1,~M_2,~ M_3)^T = \sqrt{3} U_\omega (y v,~ y' v', ~ y'' v'')^T
\label{Mi-def}
\ee
where 
\begin{equation}
U_\omega = {1\over \sqrt{3}}
\begin{pmatrix} 
1     & 1  & 1\\
1     & \omega & \omega^2\\
1      & \omega^2  &\omega
\end{pmatrix}   
\end{equation}  
is the magic matrix. 
The matrix $U_\omega$ is unitary, $U_\omega U_\omega^{\dagger} = I$ 
and the inverse $U_\omega^{-1} =  U_\omega^{\dagger} = U_\omega^*$, can be obtained 
by substitution $\omega \leftrightarrow \omega^2$.   

Then the condition
\be
y v \approx \omega^2 y' v' \approx  \omega y '' v''
\label{eq:condhier}
\ee
leads to the hierarchy of masses:
$$
m_3 = 3 y v  \gg m_{2}, ~ m_{1}. 
$$
The  hierarchy  between the lighter states, $m_{2} \gg m_{1}$,  can be obtained
by special tuning of relations in Eq.~(\ref{eq:condhier})
or introduction of  other contributions, e.g.  from the linear seesaw terms. 
 
In general,  it is difficult to get correlation (\ref{eq:condhier})
between  VEVs and Yukawa couplings. The simplest realization would be
two separate equalities:
$$
y \approx y'  \approx  y'' ,
$$
as a result of additional symmetry, or the fact that
singlets originate from breaking of representation {\bf 3}  and 
$$
v \approx \omega^2 v' \approx \omega v'',
$$
as a consequence of special forms of the potential. \\

In Eq.~(\ref{diag-off}) the off-diagonal matrix equals 
\begin{equation}
M_{33}^{\xi} = h 
\begin{pmatrix}
0   &  u_3  &  u_2\\
...      & 0   &  u_1 \\
...     & ...  & 0 
\end{pmatrix}. 
\label{zee-wolf}
\end{equation}
It has the form of the Zee-Wolfenstein 
mass matrix~\cite{Zee:1980ai, Wolfenstein:1980sy} with  the eigenvalues 
satisfying the equality $m_1 + m_2 + m_3 = 0$. 
For arbitrary  values of $u_i$, the  
diagonalization matrix and eigenvalues of $M_{33}^{\xi}$  
are presented in Appendix.~\ref{diag:off}, 
and here we consider some special cases.  

If $u_1 = u_2$,  the matrix  can be diagonalized by maximal 2-3 rotation  
and vanishing $\theta_{13}$. Then for the 1-2 mixing we obtain  
$\sin^2 2\theta_{12} = 8u_1^2/(8u_1^2 + u_2^2)$.  
If  all  VEVs are equal, $u_1 = u_2 = u_3$, then  $\sin^2 2\theta_{12} = 8/9$, 
which corresponds to the TBM mixing. 

Let us consider a special case of the total matrix $M_{33}$ (\ref{diag-off}), 
which will appear  
often in our constructions. If  $u_1 = u_2 = u_3 \equiv u$ 
and also the  diagonal elements are  
all equal,  we have 
\begin{equation}
M_{33}^{special} =
\begin{pmatrix}
\mu  & \beta &  \beta\\
...      & \mu   & \beta    \\
...     & ... &  \mu
\end{pmatrix},
\label{special}
\end{equation}
where $\beta \equiv h u$~\cite{Hirsch:2007kh}. 
Such a matrix can be obtained when $\varphi'$ and $\varphi''$ 
do not contribute to masses, or when the bare mass terms or flavonless 
operators dominate. 
This matrix having  the TBM form is diagonalized by 
$U_{TBM}$,  see Appendix.~\ref{App:mixings}.  The eigenvalues equal
\be
\lambda_1 = \lambda_3 = \mu - \beta, ~~~~ \lambda_2 = \mu + 2 \beta,  
\label{eigenvvv}
\ee
of which two are equal and the sum is given by  
$$
\sum_i \lambda_i = 3 \mu . 
$$
For $\mu = 0$ we have  $\sum_i \lambda_i = 0$,  and therefore  
$M_{33}^{special}(\mu = 0)$ can not be used 
to describe masses of quarks or charged leptons.

Degeneracy of the two eigenstates  implies  that 
the diagonalization matrix is not unique, and actually  
there is  infinite ambiguity in diagonalization
related to rotation in the space of equal eigenstates. 
In particular,  the matrix  (\ref{special})
can be {\it unitarily} diagonalized by the magic matrix as\footnote{
Note that in terms of the  SM components, the Dirac mass terms read  
$$
(l^c_L)^T C m_D  l_L + l_L^T m_D C l^c_L  + h.c.,
$$
where $l_L$ and  $(l^c)_L$  are the left and the right components, 
and $m_D$ is a symmetric matrix. Then the unitary diagonalization 
$$
U_\omega^* m_D U_\omega = {diag}(m_1,\,m_2,\,m_3),
$$
would imply that $l_L$ and $(l^c)_L$ transform differently,
but $l_L$ and $l_R$ transform in the same way.}  
$$
U_\omega^* M_{33}^{special} U_\omega. 
$$
The diagonalization gives 
$\lambda_1 = \mu + 2 \beta$ and 
$\lambda_2 = \lambda_3 = \mu - \beta$, which coincides with  Eq.~(\ref{eigenvvv}) 
up to permutation.

The ambiguity is removed when the degeneracy of eigenvalues is broken. 
Let us consider correction  matrices  of  the form  
\begin{equation}
\Delta M = M_0 
\begin{pmatrix}
1  & 0 &  0\\
0      & 0   & 1    \\
0     & 1  &  0
\end{pmatrix}~~~\text{or}~~~
\Delta M' = M_0'
\begin{pmatrix}
0  & 0 &  1 \\
1   & 0   &  0  \\
0     & 1  & 0
\end{pmatrix}, 
\label{perturbms}
\end{equation}
of which both can arise from a $Z_3$ symmetry, with different charge assignments. 
The former symmetric matrix fixes  the diagonalization matrix to $U_{TBM}$, whereas the latter one 
fixes it to $U_{\omega}$.  Note that $U_\omega$ can only be obtained from non-symmetric corrections, such as $\Delta M'$.

It is easy to get maximal mixing from  $M_{33}$, 
which will be often needed for our constructions below. 
For instance,  the maximal 2-3 mixing can be obtained if 
$\langle \xi \rangle  = (u_1, 0, 0)$ and  $y' v' = y'' v''$, or  $\vec{y} = 0$.    
For maximal 1-3 mixing one needs $\langle \xi \rangle  = (0, u_2, 0)$ and 
$y' v'  = \omega y'' v''$.  

In general,  by varying $y v$,  $y' v'$,   $y'' v''$  as well as 
$h u_i$, one can obtain an arbitrary symmetric matrix for $M_{33}$. \\

2. $F_1 \sim {\bf 3}$ and $\vec{F}_2  \equiv (F_2, F_2', F_2'') 
\sim ({\bf 1},\,{\bf 1}',\, {\bf 1}'')$. 
The most general couplings  with flavons  
\begin{equation}
\label{mass:form31}
\vec{h} ~(F_1 \xi)_{F_2} \vec{F}_2  \equiv 
h (F_1 \xi) F_2  +  h' (F_1 \xi)''F_2'  + h'' (F_1 \xi)' F_2''  
\end{equation}
generate the mass term $F_1 M_{31} F_2^T$ with 
\begin{equation}
M_{31} =
\begin{pmatrix}
h u_1  & h' u_1  &  h'' u_1\\
h u_2      &~ h' u_2 \omega   & ~ h'' u_2 \omega^2    \\
h u_3     & ~h'  u_3  \omega^2  & ~ h''  u_3 \omega
\end{pmatrix}.  
\label{eq:m1313}
\end{equation}
Obviously, only flavon triplets contribute to $M_{31}$~\cite{Ma:2004zv,Ma:2006vq}. This matrix can be represented in the form 
\be 
M_{31} = \sqrt{3} D_u U_\omega D_h~, 
\label{eq:m31}
\ee
where $D_u \equiv diag(u_1, u_2, u_3)$  
and $D_h \equiv diag(h, h', h'')$.

Additional symmetries related to transformations of components of  
$\vec F_2$ may lead, e.g. to $h = h' = h''$, thus 
further reducing the  number of free parameters in  $D_h$.  
If  two or all three fermions 
in $\vec F_{2}$ have the same symmetry assignments, including charges 
under $A_4$, e.g.  $\vec F_2 \sim ({\bf 1}, {\bf 1'},{\bf 1'}) $, 
we would obtain a singular 
mass matrix $M_{31}$ with two or three columns proportional to each other. 

In general, the matrix (\ref{eq:m1313}) is non-symmetric. It can be made symmetric
if $D_h \propto D_u$ (or equivalently, $u_i \propto h^i$), which can be obtained 
when $h^i$ themselves are given by VEVs of some new fields. \\

3. $\vec{F}_1 \equiv (F_1, ~ F_1', ~ F_1'') \sim  ({\bf 1},\,{\bf 1}',\, {\bf 1}'')$, 
$\vec{F}_2 \sim  ({\bf 1},\,{\bf 1}',\, {\bf 1}'')$.  
For the same set of the fermionic singlets, {\it i.e.}  
$\vec{F}_2  = \vec{F}_1$,   the most general coupling with flavons  
$\varphi, \varphi', \varphi''$ is given by 
\begin{eqnarray} 
& \, & m_0 F_1 F_1 + 2 \tilde{m}_0 F'_1 F''_1 
+ (y_{11}F_1 F_1 +  2 y_{23} F_1' F_1'') \varphi 
+ (y_{22} F_1' F_1' + 2 y_{13} F_1 F_1'') \varphi' 
\nonumber\\ 
& + & (2 y_{12} F_1 F_1' +  y_{33} F_1'' F_1'') \varphi'' + h.c.   
\end{eqnarray} 
It leads to the mass terms $\vec{F}_1 M_{11} \vec{F}_1^T $, where   
\begin{equation}
\label{mass:form11}
M_{11} = 
\begin{pmatrix}
m_0 + y_{11} v  & y_{12} v''  &   y_{13} v'\\
...  & y_{22} v'    & \tilde{m}_0 + y_{23} v \\
...   & ...  &  y_{33} v'' 
\end{pmatrix}. 
\end{equation} 
For all free parameters, $M_{11}$ is an arbitrary symmetric matrix.  
One can get from Eq.~(\ref{mass:form11}) maximal mixing or 
the TBM mixing immediately under the conditions 
\be
y_{22} v' = y_{33} v'',  ~~~~~ y_{12} y_{22} = y_{13}y_{33}, ~~~~
(m_0 - \tilde{m}_0)  + (y_{11} -  y_{23}) v + (y_{12} - y_{33})v'' = 0.  
\label{tbm11}
\ee
They are satisfied if, e.g. $m_0 = \tilde{m}_0$,  while all the couplings and 
all the VEVs are equal, respectively.  

If the bare mass term  
dominates, it produces the structure with maximal 2-3 mixing 
which can be used to construct the TBM or BM mixing. 

If all the couplings in Eq.~(\ref{mass:form11}) are equal 
and the bare mass terms are absent being forbidden by additional symmetry, 
the matrix $M_{11}$ acquires a form 
\begin{equation}
\label{omega:ml}
M_\omega = 
\begin{pmatrix}
a    & c  & b\\
c    &b  & a\\
b    &a  & c
\end{pmatrix}, 
\end{equation} 
where $a = h v$, $b = h v'$ and $c = h v''$. 
The matrix $M_\omega$  is diagonalized 
by the magic matrix $U_\omega$ as $U_\omega^T M_\omega U_\omega$. 
The eigenvalues of $M_\omega$  in this case equal 
$$
(m_1, m_2, m_3)^T = \sqrt{3} U_\omega^* (a,b,c)^T, 
$$
and  there is no mass (eigenvalue) degeneracy 
provided that none of $\alpha,~\omega \alpha$, 
and $\omega^2\alpha$  is real, where 
$\alpha \equiv  ab^*+bc^*+ca^*$.   
Notice that Eq.~\eqref{omega:ml} is the  only form 
of  symmetric  matrix 
which  produces the mixing $U_l = U_\omega^*$  
\footnote{If $m_D$ is non-symmetric, which can happen in the presence of additional  
$120_H$ or $45_H$ Higgs multiplets, other forms  
are allowed to generate $U_\omega$.}.

\vspace{0.15cm}

Thus,  $A_4$ with additional conditions ({\it i.e.}  equalities of couplings 
and VEVs, which in turn imply 
some additional symmetries) allows to get $U_{TBM}$ or $U_\omega$. 
The latter appears from various interactions, and 
together with additional maximal 1-3 rotation, can again 
lead to $U_{TBM}$. In this sense 
$A_4$ has features which are close to the TBM mixing. \\

For $16_F \sim {\bf 3}$, the Dirac mass matrices are of $M_{33}$ form:  
$m_D = M_{33}^{(D)}$.
With certain restrictions they can be further of $M_{33}^{\varphi}$ or $M_{33}^{\xi}$ type.  
If $S \sim {\bf 3}$, all other matrices also  have the
form $M_{33}$: $M_D = M_{33}^{(P)}$, $M_S = M_{33}^{(S)}$
although  values of elements of these matrices  being functions of
different couplings $(y, h)$ and  VEVs 
$(v, u)$, are different. This is marked by different superscripts. 
Here also additional restrictions can select  $M_{33}^{diag}$ or $M_{33}^{\xi}$. 
All these matrices are symmetric. 
The mass matrix of light neutrinos equals
\be
m_\nu =  M_{33}^{(D)} \left[M_{33}^{(P)T}\right]^{-1} \cdot M_{33}^{(S)} \cdot 
\left[M_{33}^{(P)}\right]^{-1} M_{33}^{(D)}.  
\label{eq:matrix3-3}
\ee

For $S \sim ({\bf 1},\,{\bf 1}',\, {\bf 1}'')$  the portal and singlet
mass matrices have different structures:
$M_D = M_{31}^{(P)}$ and $M_S = M_{11}^{(S)}$. Correspondingly,
the light neutrino mass matrix is given by 
$$
m_\nu = M_{33}^{(D)} (M_{31}^{(P)T})^{-1}
M_{11}^{(S)}  \left[ M_{31}^{(P)}\right]^{-1}  M_{33}^{(D)}, 
$$
and now $M_D$ is usually not symmetric. 
Using representation (\ref{eq:m31}) we have
\be
m_\nu = {1\over 3} \, M_{33}^{(D)} D_u^{-1} \cdot U_\omega^* \cdot (D_h^{-1}
M_{11}^{(S)} D_h^{-1}) \cdot U_\omega^* \cdot D_u^{-1}  M_{33}^{(D)}. 
\label{eq:matrix3-1}
\ee 
The screening 
can arise from $ M_{33}^{(D)} (M_{33}^{(P)T})^{-1}$ in Eq.~\eqref{eq:matrix3-3}, 
or partial screening from $M_{33}^{(D)} D_u^{-1}$  in Eq.~\eqref{eq:matrix3-1}, 
depending on the $A_4$ charges of fermion singlets. 

Including also the possibility of $16_F \sim ({\bf 1},~ {\bf 1'},~ {\bf 1''})$   
we obtain four different combinations of mass matrices shown 
in the Table.~\ref{A4flavor:structure}. 
In what follows we mostly focus on the first two possibilities 
with  $16_F \sim {\bf 3}$ and  $S$ transforming as  
either ${\bf 3}$ or $({\bf 1},\,{\bf 1}',\, {\bf 1}'')$.

\begin{table}[htbp] 
\center
\begin{tabular}{|c|c||c|c|c|c|c|} \hline
~~$16_F$ under $A_4$~~ & ~~$S$  under $A_4$~~ & ~~Dirac $m_D$~~& ~~portal $M_D$~~ 
&~~singlet $M_S$~~  \\\hline
{\bf 3}  & {\bf 3}    &  $M^{(D)}_{33}$ &  $M^{(P)}_{33}$ & $M^{(S)}_{33}$ \\\hline
{\bf 3}     & ({\bf 1},\,{\bf 1}$'$,\,{\bf 1}$''$)  & 
$M^{(D)}_{33}$  &  $M_{31}^{(P)}$  &  $M^{(S)}_{11}$ \\\hline
({\bf 1},\,{\bf 1}$'$,\,{\bf 1}$''$)     & {\bf 3}  &  
$M^{(D)}_{11}$  & $M_{13}^{(P)}$  & $M^{(S)}_{33}$ \\\hline
({\bf 1},\,{\bf 1}$'$,\,{\bf 1}$''$)&({\bf 1},\,{\bf 1}$'$,\,{\bf 1}$''$)&$M^{(D)}_{11}$  &  
$M^{(P)}_{11}$  & $M^{(S)}_{11}$\\\hline
\end{tabular}
\caption{Combinations of mass matrices for various $A_4$ assignments.}
\label{A4flavor:structure}
\end{table}

In general,  any structure of $m_\nu$ can be obtained. 
So, additional restrictions are needed  
in order to produce specific flavor structures. 
Further structuring can be achieved  
if  not all possible flavons are introduced when 
some couplings and VEVs  of flavons are forbidden (zero) 
or are related to each other. 
The latter implies an additional symmetry, 
such as $Z_2$ or $Z_4$.  
If more than three fermionic singlets $S$ exist we can obtain
more flavor structures assigning $S$ in triplet and
singlet representations of $A_4$: $S \sim {\bf 3} + ({\bf 1},\,{\bf 1}',\, {\bf 1}'')$. 
This extension will be considered in some details in sect.~\ref{chapt:exte}.

Finally, 
in the case of $16_F$ being singlets of $A_4$,  
$m_D = M^{(D)}_{11}$ in the form of Eq.~\eqref{mass:form11}, 
and some new possibilities appear: 

1. As we have marked already, equal Yukawa couplings 
(without  bare masses) lead to the Dirac mass matrix
of the form  Eq.~\eqref{omega:ml}. The later is diagonalized by 
$U_\omega$. Then maximal 1-3 mixing can follow from the portal and hidden sector interactions to produce the TBM mixing.

2. If $S$ are transforming as  triplets of $A_4$,  
partial screening, {\it i.e.} $m_D/M_D \propto U_\omega$, can  be
obtained  since in this case 
$M_{13}^{(P)} \sim M_{31}^T \sim  D_h U_\omega D_u$. 
However, such a  construction usually corresponds to a diagonal
$m_D$, which is difficult to generate from $m_D = M_{11}$.

\subsection{Factorization of mixing matrix}  
\label{TBM:Factorizing}

In general, the lepton mixing matrix can be constructed as 
$$
U_{PMNS} = V_0 \times U_1 \times U_2 \times ..., 
$$
where  different factors come from diagonalization of different mass matrices involved. 
In our approach we take $V_0 \sim V_{CKM}^\dagger$. 
This common part for the quarks and leptons is due to 
the SO(10) GUT and  the CKM physics. The rest, 
$U_0 \equiv U_{PMNS^0} = U_1 \times U_2 ...$,  reproduces  
$U_{TBM}$ or $U_{BM}$, {\it i.e.} matrices of special type 
dictated by the flavor symmetry.  Here we focus on construction 
of $U_0$  from the double seesaw mechanism, and discuss its TBM or BM forms.  
The CKM mixing, the non-zero 1-3 mixing, as well as other deviations from the TBM mixing, 
are probably related to each other and have some other 
origin, which we refer to as the CKM new physics. 
In view of small CKM mixing, $U_0$  can be considered as  the zero-order 
approximation of the PMNS mixing, $U_{PMNS^0}$. 

Flavor features which lead to TBM matrix have been extensively 
studied before (see \cite{Lin:2008aj, Altarelli:2010gt} and 
references therein). The real TBM matrix can be factorized as  
$$
U_{TBM} = {\rm diag} (1, \omega^2, \omega) U_\omega U_{13}^{max T} 
{\rm diag}(1, 1, i). 
$$
This factorization is not unique: in general any product of 
$U_{\omega}$ (also with permuted rows and columns) and 2-state maximal mixings,   
$U_{\omega} U_{12}^{max}$, $U_{\omega} U_{13}^{max}$, or $U_{\omega} U_{23}^{max}$,  leads to
the TBM matrix with permuted rows and columns up to rephasing. 
This, however,  does not affect physics: flavor content of the eigenstates 
with certain eigenvalues 
is  the same for all the cases. \\

It is straightforward to generalize the factorization 
via ${U}_\omega$ and ${U}_{13}^{max T}$  to 
$$  
\tilde{U}_\omega = D_\alpha U_\omega U_A, ~~~~~
\tilde{U}_{13}^{max T} =  U_A^{\dagger} U_{13}^{max T} D_\beta, 
$$
where $U_A$ is arbitrary unitary matrix, and  $D_\alpha$ is arbitrary 
diagonal phase  matrix and the  phases can be absorbed into  the wave 
functions of the charged leptons.  
The matrix  $D_\beta$  contributes 
to the Majorana phases of neutrinos. In the generalized form, 
one may use $U_A = (U_{12}^{max T} U_{13}^{max})^\dagger$, so that 
$U_{PMNS} = \tilde{U}_{\omega} U_{12}^{max T}$ and one of the factors is maximal 1-2 
rotation instead of 1-3 rotation. However,  now 
$\tilde{U}_{\omega} = D_\alpha  U_{\omega} U_{13}^{max T}  U_{12}^{max}$, which is not reduced to
 $U_{\omega}$ with permuted columns (rows),  and so it is difficult to obtain  
$U_\omega$ as a result of symmetry. 

Another possible factorization of $U_{TBM}$, without using 
the magic matrix $U_\omega$ directly, is  $U_{23}^{max} U_{12}(\theta_{TBM})$, 
where $\sin^2 \theta_{TBM} = 1/3$. However, 
mass matrices diagonalized by $U_{12}(\theta_{TBM}) $ do not naturally 
appear from $A_4$ symmetry. \\

According to the double seesaw, there are three  
mass matrices relevant for generation of the lepton mixing: 
\be
m_D,\, \, \, m_D  M_D^{T\,-1}, \, \, \, M_S,  
\label{eq:threemat}
\ee
where the first matrix is the mass matrix of  charged leptons, the second one is  the screening 
factor and the third one is the mass matrix 
of singlet fermions. In general all these matrices provide contributions to the lepton mixing matrix. 
Different mass matrices in Eq.~(\ref{eq:threemat})
can be responsible for generation of different factors in the TBM matrix.  
Let us mention some possibilities mostly related to $U_\omega$ and  $ U_{13}^{max}$,  
which will be realized in specific schemes in sect.~\ref{sect:trip}. 

1.   $U_\omega$ may  follow from the charged leptons, and diagonalize the matrix $m_D$,  
whereas $U_{13}^{max}$ comes from the neutrino mass matrix,   
so that
\begin{equation}
m_\nu=
\begin{pmatrix}
x     & 0  &  z\\
0      &y  & 0\\
z      &0  &x
\end{pmatrix}.  
\label{eq:unu1}
\end{equation}
In the  SO(10)  framework $U_l = U_\omega^*$ implies that 
$m_l = m_D = M_\omega$  defined in Eq.~(\ref{omega:ml}).

2. $U_l = {\mathbbm 1}$, which means that $m_D$ is diagonal,  
 $ U_\omega$ follows from  the portal factor:  
$m_D  M_D^{T\,-1} \propto U_\omega^*$, and    
$U_{13}^{max}$ originates from  $M_S$.

3. One extreme possibility is when $U_{TBM}$ comes completely from charged 
leptons, $U_l = (U_\text{TBM})^\dagger$.  
The  neutrino diagonalization  matrix should be the unit matrix.

4. Another extreme is when the whole $U_{TBM}$ mixing is generated by  
$M_S$ ({\it i.e.} by the hidden sector):  
$m_D = diag$, $m_D^T M_D^{-1} \propto {\mathbbm 1}$,  $M_{S} = M_{TBM}$. 
In this case $m_\nu \propto M_S$.

\subsection{Additional symmetries and VEV alignments}

In what follows  we construct  schemes for masses and mixing
using mass matrices elaborated in this section  as building blocks.  
It is easy to show that in the formulated framework  
without additional constraints any 
mass matrix of the light neutrinos can be obtained. 
Therefore to get certain flavor structures  we will use 

- minimal possible number flavon fields;  

- minimal number of (different) couplings;  

- the most symmetric situations 
when as many as possible couplings are either zero or equal. 
The latter implies certain extended symmetries and for definiteness    
we will use additional symmetry $Z_4$.

Existence of different energy scales in this framework,  
$m_D \ll M_D \ll M_S$, allows to realize the idea that different 
symmetries are realized at different scales with larger symmetry at higher mass scale. 
Higher scale has larger symmetry, which is then broken explicitly or spontaneously at 
lower scale. Thus,  $M_S$ may have additional symmetry which is not realized  
in portal and visible interactions. Feedback of this breaking onto higher scale 
physics is expected to be suppressed by factors $M_D/M_S \sim 10^{-2}$.    
In particular, due to additional symmetries $M_S$  can have the TBM form. 
Although the symmetry is not realized at the portal scale, the corrections to TBM mixing 
can be of the order $10^{-2}$.  See also \cite{Ludl:2015tha} for some realizations of this idea.

We formulate conditions on couplings and VEVs which should be satisfied to get the required 
fermion masses and mixing. We provide some hints of how the conditions can be 
realized and where they can originate from.

Mechanisms have been elaborated
which allow to get  relations between VEVs
of multiplets. In particular  for triplets the alignments $v(1, 1, 1)$, or
$v(0, 1, 1)$, $v(0, 0, 1)$ can be obtained, as a consequence
of symmetry of the potential~\cite{Barbieri:1999km, Feruglio:2009iu, Holthausen:2011vd}. 
Usually  such potentials have several degenerate vacua, and so 
some mechanisms of selection  should exist.   
Construction of models which realize the conditions is beyond the scope of this paper.

In various cases we need also correlations of  VEVs of singlet  and triplet flavons:  
\be 
h(u_1, u_2, u_3)^T = \sqrt{3} U_\omega (y v,~ y' v', ~ y'' v'')^T.  
\label{uv-correlation}
\ee 
Multiplied by $U_\omega^*$, the relation  can be inverted to:  
$$
(y v,~ y' v', ~ y'' v'')^T = h  U_\omega^* (u_1, u_2, u_3)^T/\sqrt{3}. 
$$
For equal constants $y' =  y''  = y$, it can be written as 
\be
(v,~ v', ~ v'')^T =  a   U_\omega^*   (u_1, u_2, u_3)^T, 
\label{uv-correlation1}
\ee
where $a = h/\sqrt{3}y$. 

To reach the correlation (\ref{uv-correlation})   
we introduce a {\it dimensionless} auxiliary field $A$ 
with the VEVs $\langle A \rangle = (a,\,a,\,a)/\sqrt{3}$. 
Then the equality (\ref{uv-correlation})  is satisfied automatically if 
\begin{eqnarray}
v =  \langle  (\,A\cdot \xi \,) \rangle\,, ~~~~
v' =   \langle  (\, A\cdot \xi\,)' \rangle\,, ~~~~
v'' =   \langle  (\,A\cdot \xi\,)'' \rangle\, . 
\label{aux:cancel}
\end{eqnarray}
For non-equal constants $(y,\,y',\,y'')$ we should substitute 
$v' \rightarrow v' y'/y$ and $v'' \rightarrow v''   y''/y$ in the 
left-hand sides of these equalities. 
In turn,  the relation  
(\ref{aux:cancel}) can be induced e.g.  by scalar potentials of the form 
$$
V(\varphi)_\text{eff} \propto -[\varphi - (A\cdot \xi) ]^4. 
$$
Consequently,  $A \cdot \xi$  should have the same $Z_4$ charges   
as $\varphi$.

\section{Schemes with $S$ transforming as a triplet of $A_4$}
\label{sect:trip}

For single complete set of the flavons $\xi \sim  {\bf 3}$, 
$\vec{\varphi} \sim  \vec{\bf 1}$ 
the most general $A_4$ symmetric Yukawa Lagrangian (at the lowest order) is 
\begin{eqnarray}
{\mathcal L_{D}} &=&  \frac{10_H}{\Lambda_f} \left[
\vec{y}_{10}\, (16_F \cdot 16_F)_{\varphi}  \vec{\varphi} + 
h_{10}  (16_F\cdot 16_F \cdot \xi)  \right] + 
y_{10}^b\, (16_F \cdot 16_F) 10_H, \nonumber\\
{\mathcal L_{NS}} &=&   \frac{\overline{16}_H}{\Lambda_f}  
\left[\vec{y}_{16}\, (16_F \cdot S)_{\varphi}  \vec{\varphi} +  
h_{16}  (16_F\cdot S \cdot \xi) \right] +   y_{16}^b\, (16_F \cdot S) \overline{16}_H, 
\nonumber\\
{\mathcal L_{S}} & = & \vec{y}_{1}\,  (S \cdot S)_{\varphi} \vec{\varphi}   
+ h_{1} \,(S \cdot S \cdot {\xi}) + M^0_S (S \cdot S)  \, .  
\label{s-triplet}
\end{eqnarray}
Here 
$$
\vec{y}_{10}  (16_F \cdot 16_F)_{\varphi} \vec{\varphi}  \equiv 
y_{10}\, (16_F \cdot 16_F) {\varphi} + y'_{10}
\,(16_F\cdot 16_F)''   {\varphi'} + y''_{10}  (16_F\cdot 16_F)' {\varphi''},  
$$
{\it etc.}, with general definitions of these combinations given in (\ref{notation1}). 
The terms  ${\mathcal L_{S}}$ contain renormalizable couplings. 
The symmetry allows also to introduce renormalizable flavonless 
couplings of $16_F$   and $S$.  
If several sets of flavons $\vec{\varphi}_a$ and $\xi_b$ exist, 
for each of them one should introduce  
interactions similar to those in eq. (\ref{s-triplet}).

All the mass matrices ($m_D$,  $M_D$ and $M_S$) generated by this Lagrangian  
are of the $M_{33}$ type 
plus matrices proportional to unit matrix. (The latter can be forbidden by additional symmetry.) 
If the same flavons contribute to all three mass matrices, 
these matrices  are not completely independent
being functions of the same set of VEVs. 
On the other hand, sets of coupling constants for three types of interactions, 
$\vec{y}_i \equiv (y_i, ~y_i', ~y_i'')$ and $ h_i$ $(i = 10, 16, 1)$,  can be totally different. 
With this one can get arbitrary matrices $m_D$,   $M_D$  and $M_S$, 
and consequently,  arbitrary symmetric mass matrix of the light neutrinos.  

In what follows we will study the situations that some interactions 
are forbidden by  additional symmetry $Z_4$.   
One possibility is to keep minimal number of couplings in visible and portal sectors,
and to obtain all the rest from the hidden sector. 
Among four types of couplings in the visible sector and portal interactions, 
$\vec{y}_{10}$, ${h}_{10}$, $\vec{y}_{16}$ and $h_{16}$, 
at most two can be zero (one for $10_H$ and one for $\overline{16}_H$ if we do not take into account 
low-dimension interaction couplings $y^b$).  
Next complication is  to assume that some couplings among $\vec{y}_{10}$ 
and $\vec{y}_{16}$  are zero, which would require more symmetries.   
The couplings  of the  hidden sector $\vec{y}_{1}$  and ${h}_{1}$,  are also restricted 
by  $Z_4$.  Thus, we have four possibilities 
determined by  representations of flavons that contribute 
to the corresponding couplings/mass terms:  
${h}_{10} = {h}_{16} = 0$, which we call the singlet-singlet scheme; 
$\vec{y}_{10} = \vec{y}_{16} = 0$ - the  triplet-triplet scheme; 
${h}_{10} = \vec{y}_{16} = 0$ - the singlet-triplet scheme; and 
$\vec{y}_{10} = {h}_{16} = 0$ - the triplet-singlet scheme. 
We consider these possibilities in order and then present schemes 
in which both singlets and triplets contribute to the same mass terms.

\subsection{Singlet - singlet flavon scheme. } \label{sect:ssss}

\begin{figure}[t]
\centering\includegraphics[width=4.0in]{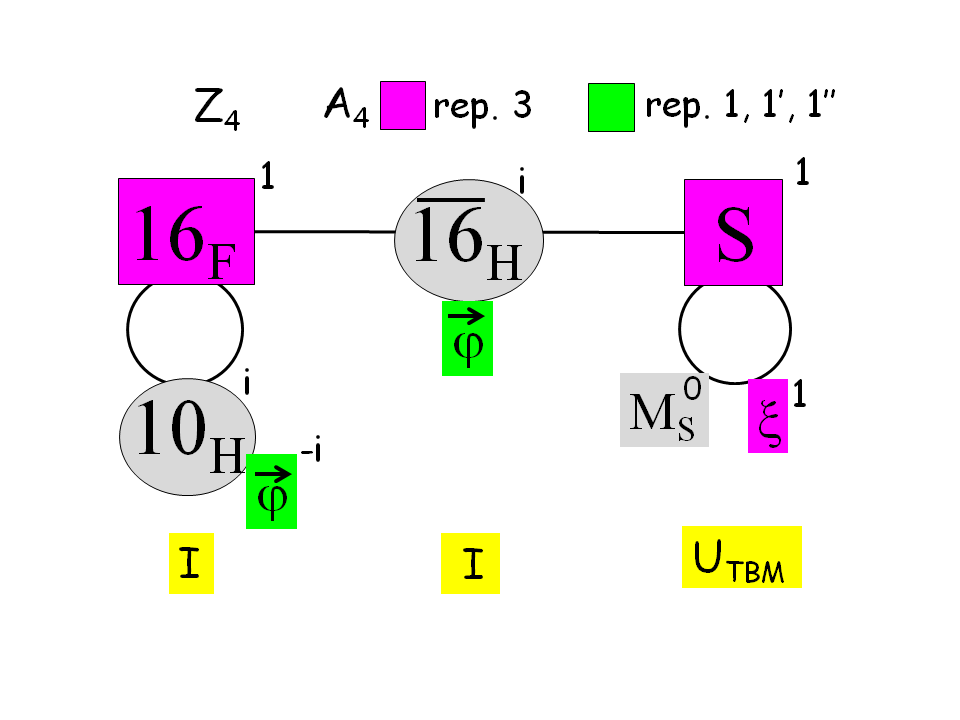}
\vskip -1.5cm
\caption{Graphic representation of the singlet - singlet flavon scheme.
Numbers at the field notations ({\it i.e.} the superscript of each particle content) are the corresponding $Z_4$ charges. In the bottom 
we show rotation matrices generated by the corresponding  interactions. }
\label{fig:ss-scheme}
\end{figure}

The scheme with $h_{10} = h_{16} = 0$  allows to 
realize  the following scenario: 
$$
U_l = {\mathbbm 1}, ~~~~~  m_D M_D^{-1} = {\mathbbm 1}, ~~~~U_S = U_{TBM},  
$$ 
where $U_l$ diagonalizes $m_D$, and $U_S$ diagonalizes $M_S$. 
Here the TBM mixing  comes from solely the hidden sector\footnote{Generation of lepton 
mixing in the hidden sector has been 
considered in \cite{Ludl:2015tha}. The difference is that here we use common 
non-abelian symmetry in visible and hidden sectors.}.  
Only singlet flavons and possibly flavonless operators  
generate  visible and portal masses. 
The triplet flavon  couplings, $h_{10}$ and $h_{16}$, are  forbidden by  
additional $Z_4$ symmetry. 
We introduce flavon singlets $\vec{\varphi}$, which operate
in the visible sector and portal terms.  One flavon  triplet $\xi$  of $A_4$ act in the hidden sector. 
For symmetry assignment under $Z_4$ we use
\be
\left\{16_F, S| 10_H, \overline{16}_H | \vec \varphi,  \xi \right\} 
\sim \left\{1, 1 | i, i | -i,  1  \right\}.
\label{new-assmn}
\ee
Graphic representation of the scheme  is shown in Fig.~\ref{fig:ss-scheme}.

The most general  $A_4\times Z_4$ symmetric Lagrangian has the form
\begin{eqnarray}
{\mathcal L_{D}} &=&  \frac{10_H}{ \Lambda_f}  \vec{y}_{10}~ (16_F \cdot 16_F)_\varphi ~ 
\vec{\varphi},  
\nonumber\\
{\mathcal L_{NS}} &=& 
\frac{\overline{16}_{H}}{\Lambda_f} 
 \vec{y}_{16} ~(16_F \cdot S)_\varphi ~\vec{\varphi} ,  
\nonumber\\ 
{\mathcal L_{S}} &=& M_S^0\, (S \cdot S  )  \, +  
{h}_{1} \,(S \cdot S \cdot \xi)  \,.
\nonumber
\end{eqnarray}
All the  mass matrices generated by this Lagrangian are of $M_{33}$ type. 
Since triplet flavons do not appear  in the visible 
and portal sectors,  $m_D$ and  $M_D$  are of $M_{33}^{diag}$ type (\ref{33diag}) 
leading to $U_l = {\mathbbm 1}$. 
Three eigenvalues of them are arbitrary.

For $M_D$ we have an expression similar to $m_D$ with substitutions 
$\vec y_{10} \leftrightarrow \vec y_{16}$ and $10_H \leftrightarrow  \overline{16}_H$. 
Complete screening requires proportionality (correlation) of  the couplings constants 
of the Dirac and portal interactions,  $\vec{y}_{10}$ and $\vec{y}_{16}$: 
\be
y_{10} : y'_{10} : y''_{10}  = y_{16} : y'_{16} : y''_{16}. 
\label{equalconst}
\ee 
This is enough for the screening 
since the same flavon fields contribute to both mass matrices. The proportionality of Eq.~(\ref{equalconst}) 
is rather a generic requirement for several schemes, so we discuss it in some details. 
This proportionality can be obtained  in various ways as follows: 

(i)  All the coupling constants  for $10_H$ and  independently $\overline{16}_H$ are  equal  to
each other (or approximately equal): 
$$
y_{10} = y'_{10} = y''_{10}, ~~~~ y_{16} = y'_{16} = y''_{16}. 
$$ 
In this case  correlations are not needed. 
The equalities can be residuals of embedding of $A_4$  
into $SO(3)$,  so that the  singlet of $A_4$ $(1, 1', 1'')$  
steam from the triplet representation of $SO(3)$. 
The (approximate)  equality of the coupling constants is also needed to obtain  
the hierarchy of masses of quarks and charged leptons. 

(ii)  The proportionality can be a consequence of further unification of $16_F$ and $S$ 
in $27$-plets of $E_6$. 

In both cases (i) and (ii) the $Z_4$ charges of $16_F$ and $S$ should be equal, as given in Eq.~\eqref{new-assmn}.

(iii) The proportionality  can be a consequence of  symmetry with respect to 
permutation $16_F \leftrightarrow S$ and $10_H \leftrightarrow  \overline{16}_H$.  
In this case also $Z_4$ charges of $10_H$  and  $\overline{16}_H$ should be the same, 
which is satisfied in the symmetry assignment. 

Under condition (\ref{equalconst})   
the matrices $m_D$ and  $M_D$ have the same flavor structure,  
so that $m_D/M_D \propto  {\mathbbm 1}$.\\

Due to the absence
of singlet flavons,  $M_S$ should have all equal diagonal elements from the bare mass terms.
The VEV alignment of the triplet $\langle \xi \rangle = u(1, 1,  1)$  
leads to the matrix $M_S$ in the form 
$$
M_S = M_{33}^{special} + \Delta M_S, 
$$
where  the special matrix $M_{33}^{special}$ is given in  
(\ref{special}) with $\mu = M_S^0$ and $\beta = h_{1} u$.  
A perturbation matrix $\Delta M_S$ is needed, and should be of the form  $\Delta M$  in
Eq.~(\ref{perturbms}), so that the total $M_S$ has the TBM form 
and thus is diagonalized by $U_{TBM}$. The eigenvalues of $M_S$ equal 
$$
\lambda_{1,3} = M_S^0 - {h}_{1} u \pm M_0,~~~~~
\lambda_2 = M_S^0  + 2{h}_{1} u + M_0,   
$$
and there is enough freedom to get any spectrum\footnote{Notice that $M_0 \neq 0$ is important, 
otherwise  we would have two degenerate 
values $\lambda_{1,3} = M_S^0 - h_{1}u$. 
That would lead to   $m_{1} = m_{3}$ for light neutrinos, which 
contradicts observations.}. The light neutrino masses satisfy that $m_i \propto \lambda_i$.  

There are several ways to obtain the additional correction $\Delta M_S$ in the form of  $\Delta M$:

1). A correction proportional to $\Delta M$ but directly  contributing to $m_\nu$   
can come from additional fermion singlets $S,  S', \, S''$, 
as we will discuss in sect.~\ref{chapt:exte}.

2). Another modification of the scheme which would be equivalent to the 
correction $\Delta M$ (\ref{perturbms})
is to introduce flavons $\phi, \phi'$, $\phi''$  
which  modify the diagonal elements of $M_S$. 
Suppose  that $ \phi,  \phi' , \phi''$  
acquire VEVs   $(v_1, v_1', v_1'')$  which satisfy together with couplings the equality  
$y_{1} v_1 = y'_{1}v_1' = y''_{1} v_1''$. And  differently from above, $\langle \xi \rangle = (u_1,u_2,u_3)$.
Then the mass matrix of singlets can have the form 
\begin{equation} 
\label{natural:MS33}
M_S = h_1 
\begin{pmatrix}
3 y_1 v_1/h_1    &  u_3  &   u_2\\
...    & 0  &  u_1\\
...  & ... & 0
\end{pmatrix} + M_S^0\,{\mathbbm 1} . 
\end{equation}
In the case of complete screening it gives the light neutrino mass matrix:   
$m_\nu = b \cdot M_S$.  
Imposing further equality 
$u_3 \equiv u_2$ we obtain from (\ref{natural:MS33}) maximal 2-3 mixing, 
and still  enough freedom is left to fit all the  data.  
In particular,  equality $ h_1 u_1 = h_1 u_2  + 3 y_1 v_1$   
leads to  the TBM mixing  and to the mass eigenvalues  
$$
m_1 = b (M_S^0 +  h_1  u_1 - 2 h_1 u_2), ~~ 
m_2 = b (M_S^0 + h_1 u_1 + h_1 u_2), ~~~
m_3 = b (M_S^0 - h_1 u_1), 
$$
which can reproduce the required  mass spectrum.

\subsection{Triplet - triplet flavon scheme}
\label{sect:tttt}

\begin{figure}[t]
\centering\includegraphics[width=4.0in]{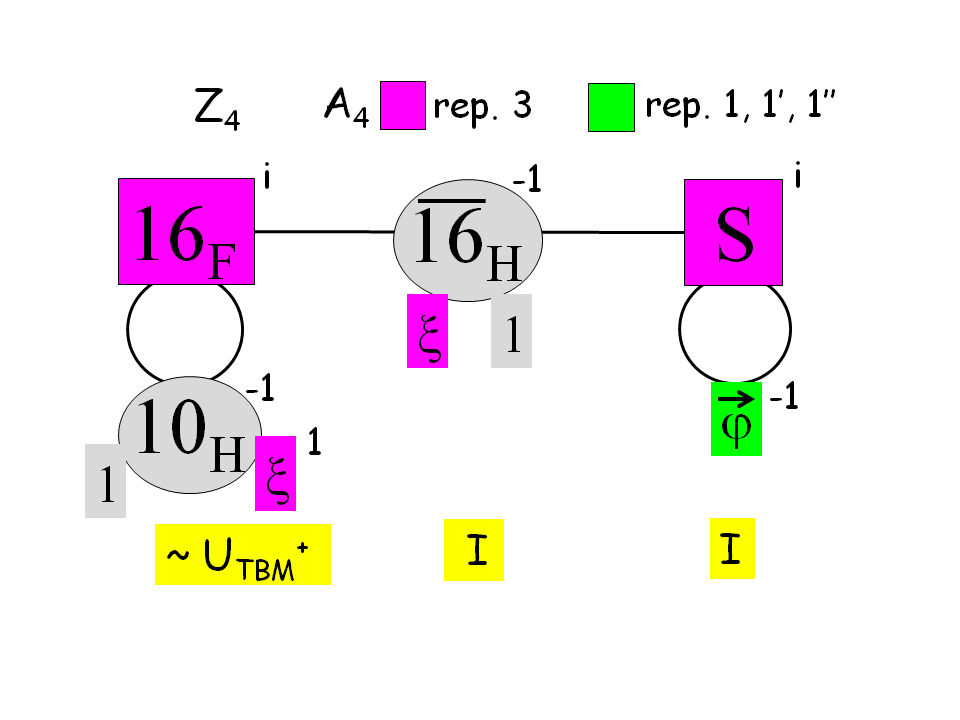}
\vskip -1.5cm
\caption{The same as in Fig.~\ref{fig:ss-scheme}, but for  the triplet - triplet flavon scheme.
Grey box with 1 corresponds to flavonless couplings. }
\label{fig:tt-scheme}
\end{figure}

The Lagrangian is given in Eq.~(\ref{s-triplet}) with 
$\vec{y}_{10}= \vec{y}_{16} = 0$ and graphic representation of the scheme 
is shown  in Fig.~\ref{fig:tt-scheme}.  
It allows to realize a scenario 
$$
U_l \approx U_{TBM}^{\dagger}  , ~~~~~ m_D/M_D \propto {\mathbbm 1}, ~~~~~ 
U_S = {\mathbbm 1}, 
$$
in which  the mixing originates from the charged leptons. 
Vanishing  singlet couplings $\vec{y}_{10}$ and  $\vec{y}_{16}$  
can be obtained by imposing the $Z_4$ assignment  
\be
\{16_F,~S |10_H,~\overline{16}_H | \vec{\varphi},~  \xi \} \sim
\{ i,~i | -1, ~ -1 | - 1, ~ 1 \}. 
\label{assignment1}
\ee  
Now  only the triplet and operators without flavons contribute 
to the visible and portal masses. Therefore 
\be  
\label{mass:trip}
m_D = m_D^0 \cdot{\mathbbm 1} + r M_{33}^{\xi}, ~~~~ 
M_D = M_D^0 \cdot {\mathbbm 1} + M_{33}^{\xi}, 
\ee
where  
$m_D^0 = y_{10}^b \langle 10_H \rangle$,  $m_D^0 = y_{16}^b \langle \overline{16}_H \rangle$  
and  $r \equiv  h_{10} \langle 10_H \rangle / h_{16} \langle \overline{16}_H \rangle$.

Diagonalizing $m_D$ we find that  masses  of charged leptons can be reproduced if 
\be
|m_D^0| = \frac{1}{3}(m_e + m_\mu + m_\tau),~~~ 
(h_{10})^2 \sum_i u_i^2 \approx \frac{1}{3} m_{\tau}^2, ~~~~
(h_{10})^3 u_1  u_2  u_3 \approx  \frac{1}{27} m_{\tau}^3. 
\label{elements}
\ee
The second and third conditions are satisfied if, e.g.  
$u_1 \approx u_2 \approx   u_3 \approx  m_{\tau}/ (3 h_{10})$. \\

One important feature of this case is that 
the matrices $m_D$ and $M_D$ in Eq.~(\ref{mass:trip}) have different eigenvalues 
but the same  mixing. 
Since off-diagonal parts of both matrices are generated by the same flavon field $\xi$, 
they are diagonalized by the same rotation 
$U_{33}$ which diagonalizes $M_{33}^{\xi}$:  
$U_{33}^T M_{33}^{\xi} U_{33} = M_{D}^{diag}$ (see Appendix.~\ref{diag:off}).  
At the same time, depending on relative size  
of diagonal and off-diagonal contributions, the matrices $m_D$ and $M_D$  
can have different hierarchies of eigenvalues.  
Consequently, $U_l = U_{33}$ and in the flavor basis the mass matrix of light 
neutrinos becomes 
\be 
\label{mass:poss2}
m_\nu = m_D^{diag} (M_D^{diag})^{-1}  U_{33}^T M_S U_{33}  (M_D^{diag})^{-1} m_D^{diag}. 
\ee
Complete screening, $m_D (M_D)^{-1} = c\cdot {\mathbbm 1}$,   
requires that  $y_{10}^b : y_{16}^b = h_{10} : h_{16}$. 
In this case we obtain from Eq.~(\ref{mass:poss2})   
\be
\label{mass:poss3}
m_\nu = c^2 U_{33}^T M_S U_{33} = c^2 U_{l}^T M_S U_{l}
\ee
in the flavor basis.

In the hidden sector  
due to  $Z_4$ symmetry given by Eq.~(\ref{assignment1})  $h_1 = 0$,  and so the 
flavon triplet does not contribute to the singlet mass matrix. 
Therefore   $M_S = M_{33}^{\varphi}$ and the neutrino matrix in Eq.~(\ref{mass:poss3})  is diagonalized 
by $U_{33}^\dagger$. Consequently,  the neutrino mixing  $U_0 = U_{33}^\dagger$. 
This result can be immediately obtained by noticing that in the original 
symmetry basis and in the presence of complete screening  the neutrino mass matrix is diagonal 
whereas the  lepton mass matrix is diagonalized by $U_l = U_{33}$.

In turn, the matrix $U_{33}$ can reproduce the TBM or BM mixing. Indeed, 
if the 1-3 mixing is generated by the  CKM physics, the 1-3 mixing,   
which originates from  $U_{33}$,  should be  zero. 
So, in terms of the standard parametrization of $U_{PMNS}$  
we have to have $U_{33} = U_{12}(\theta_{12})^T U_{23}(\theta_{23})^T $.  
This requires equality $u_2 = u_1$ in $M_{33}^\xi$ (\ref{zee-wolf})  
which leads to the maximal 1-2 mixing, $\theta_{12} = \pi/4$. Then for the 2-3 
mixing we  obtain 
$$
\tan 2\theta_{23} = \frac{2 \sqrt{2} u_1}{u_3}.  
$$
For $u_1 \gg u_3$ this equation would give nearly maximal 2-3 mixing. 
However, this is not consistent with  (\ref{elements}) which has for 
$u_2 = u_1$ two solutions: $u_3 = u_1$ and $u_3 =  (2\sqrt{2}i) u_1$. 
In the first (best) case we obtain $\theta_{23} = 35^\circ$, which can be slightly 
bigger if $u_2 \neq u_1$, but the latter corresponds to  non-zero 1-3 mixing. 
Thus, the  scheme produces the bi-large  mixing from the neutrino sector. 
Neutrino masses are given by eigenvalues of $M_{33}^{diag}$.

\subsection{Singlet -  triplet  flavon scheme}

\begin{figure}[t]
\centering\includegraphics[width=4.0in]{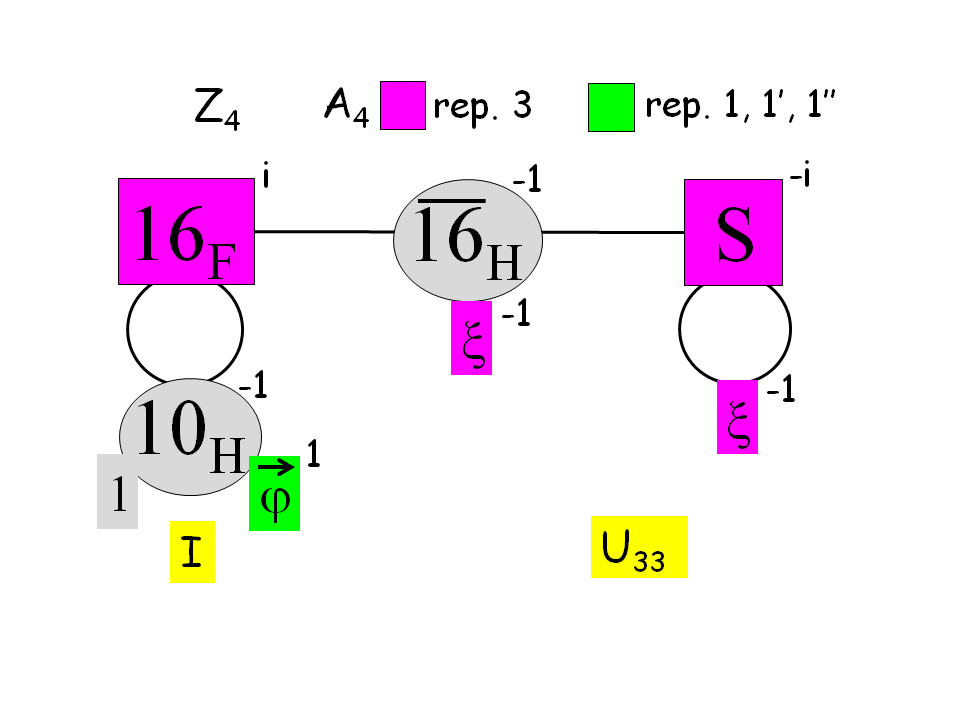}
\vskip -1.5cm
\caption{The same as in Fig.~\ref{fig:ss-scheme}, 
but for  the singlet - triplet flavon scheme. The matrix which diagonalize $m_\nu$ comes 
from a non-trivial combination of $U_{33}$ and $m^{diag}_D$ (see the text).  
}
\label{fig:ts-scheme}
\end{figure}

The Lagrangian of this scheme  is given by Eq.~(\ref{s-triplet})
with $h_{10} = \vec{y}_{16} = 0$, and the graphic representation is shown in  
Fig.~\ref{fig:ts-scheme}. Only flavon singlets contribute 
to the Dirac mass matrix and only triplet flavons to the portal mass matrix. 
Here we can realize the scenario  
$$
U_l = {\mathbbm 1}, ~~~~ M_D \propto M_S = M_{33}^{\xi}
$$
with cancellation of the portal and hidden sector matrices in the double seesaw mechanism. 
%
Vanishing $h_{10}$  and $\vec{y}_{16}$  
can be obtained  with $Z_4$ symmetry assignment  
\be
\{16_F,~S |10_H,~\overline{16}_H | \vec{\varphi},~  \xi \} \sim \{ i,~ - i | -1, ~ -1 |  1, ~ - 1 \}.
\label{z4ass1}
\ee 
The key feature is that  $16_F$ and $S$ have different $Z_4$ charges. 
The $Z_4$ symmetry  also forbids the low-dimension (flavonless) contribution 
of the mass $ M_D^0$,  so that  
$$ 
m_D = m_D^0{\mathbbm 1} +  M_{33}^{\varphi} , ~~~~ M_D =  M_{33}^{\xi}, 
$$
where $m_D^0 = y^b \langle 10_H \rangle$. 
The matrix $m_D$ has enough free parameters 
to reproduce the charged lepton masses.

With the assignment in Fig.~\ref{fig:ts-scheme}, only triplet $\xi$ contributes to the hidden sector matrix,   
so that  $M_S \propto M_{33}^{\xi}$. 
Furthermore, since the same $\xi$ contributes to both $M_S$ and $M_D$ we have 
$M_S \propto M_D$. Therefore $M_S$   
cancels with one inverse $M_D$ in the double seesaw formula.   
As a result, the light neutrino mass matrix becomes 
\be
m_\nu \propto m^{diag}_l~ (M_{33}^{\xi})^{-1}~ m^{diag}_l.  
\label{nunu}
\ee
Effectively the expression for mass  looks like a single seesaw and only the overall scale of the 
RH neutrino masses imprints information about the presence of the hidden sector. 

In spite of strong hierarchy in $m^{diag}_l$
the neutrino mass matrix in (\ref{nunu}) can be made in agreement with data due 
to the special form of $M_{33}^{\xi}$, such that its inverse matrix,  
\be
(M_{33}^{\xi})^{-1} = \frac{1}{2h u_1 u_2 u_3}
\begin{pmatrix}
- u_1^2    &  u_1 u_2   &  u_1 u_3\\
...   &  - u_2^2   &  u_2 u_3 \\
...   &  ...   & - u_3^2
\end{pmatrix}, 
\label{matinv}
\ee
has quadratic hierarchy in $u_i$.  
To show this  we take 
$m^{diag}_l = m_\tau  diag (\lambda^6, \lambda^2, 1)$, 
where $\lambda \equiv \sin\theta_C\simeq 0.22$ is the Cabibbo angle. 
Then for VEV  the hierarchy $\langle \xi \rangle = (u_1, u_2, u_3) 
= (- 1, \lambda^3, \lambda^5$) with coefficients of the order 1  
(which is even weaker than the hierarchy in $m^{diag}_l$)  
Eqs. (\ref{nunu}) and    (\ref{matinv}) lead to
$$
m_\nu \propto 
\begin{pmatrix}
- \lambda^4    &\lambda^3  &   \lambda^3\\
...   &   \lambda^2  & -\lambda^2 \\
...    &  ...  & \lambda^2
\end{pmatrix} .  
$$
This matrix produces a zero (small) $\theta_{13}$, maximal 2-3 mixing and 
large 1-2 mixing.  It only corresponds to the  normal mass hierarchy.

\subsection{Triplet -  singlet  flavon scheme}

\begin{figure}[t]
\centering\includegraphics[width=4.0in]{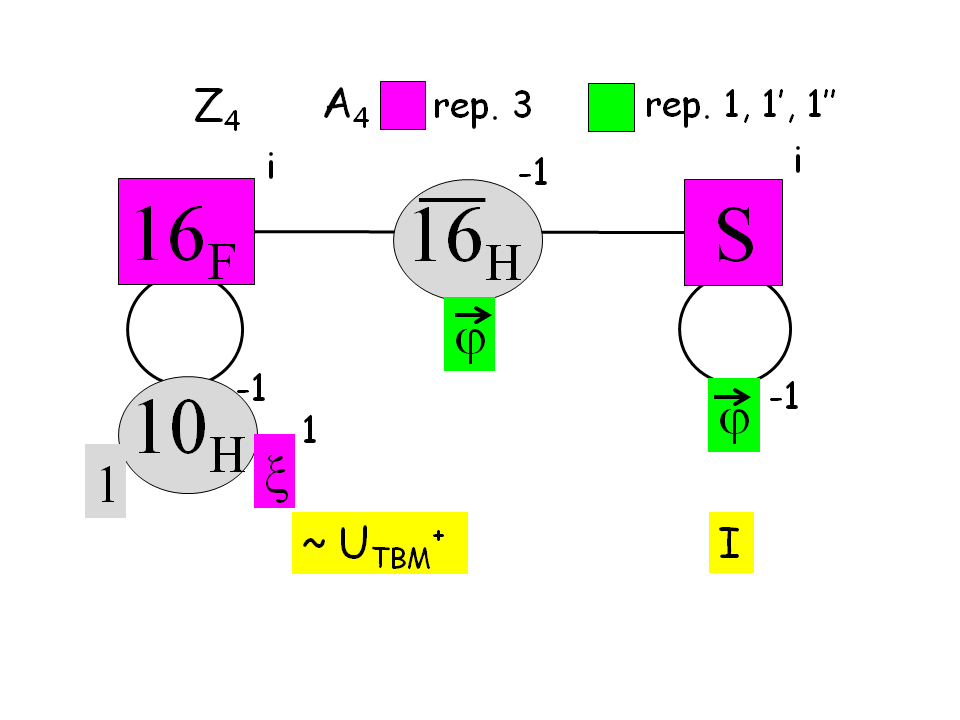}
\vskip -1.5cm
\caption{The same as in Fig.~\ref{fig:ss-scheme}, but for  
the triplet - singlet flavon scheme.}
\label{fig:st-scheme}
\end{figure}

This case  is in some sense opposite to the previous one: 
the triplet $\xi$ produces the  Dirac mass matrix,  whereas singlets generate the  portal 
mass matrix, see Fig.~\ref{fig:st-scheme}.  
The Lagrangian is given by Eq.~(\ref{s-triplet}) with 
$\vec{y}_{10}= {h}_{16} = 0$. 
Vanishing $\vec{y}_{10}$ and  ${h}_{16}$  
can be obtained  by the $Z_4$ with assignment  
$$
\{16_F, S | 10_H,  \overline{16}_H | \vec{\varphi}, \xi \} \sim  \{i,  -i | - 1, - 1 | - 1, 1 \}. 
$$ 
It  leads to
$$ 
m_D = m_D^0 \cdot{\mathbbm 1} + M_{33}^{\xi} , ~~~~  M_D = M_{33}^{\varphi}.  
$$
Symmetry allows flavonless contribution 
$m_D^0$,  whereas a similar contribution to $M_D$ is forbidden. 
Generation of the Dirac mass matrix $m_D$  is similar to the case  in sect.~\ref{sect:tttt}.

The scheme allows to realize an interesting possibility 
when mixing comes essentially from the RH rotations which diagonalize 
the neutrino Dirac mass matrix. This is possible due to the Majorana nature of neutrinos. 
If $M_D \propto \mathbbm{1}$ (only the true singlet, $\varphi$, contributes),  then 
$$ 
m_\nu \propto m_D^{diag} \cdot U_{33}^\dagger M_S U_{33}^* \cdot  m_D^{diag} 
$$
in the flavor basis, and $m_D^{diag} = m_l^{diag}$. 

Due to  $Z_4$  only $\vec \varphi$ can couple to  $S$, and consequently, $M_S$ is diagonal.  
Even in this case by selecting parameters of $U_{33}$ one can reproduce observables. 

\subsection{Scheme with mixed contributions: the TBM mixing}\label{TBM:33mixing}

Here and in the next subsection  we consider general 
situations when both triplet and singlet  
flavons interact in the visible sector. 
The model in this subsection realizes the scenario in which 
$$
U_l = U_{\omega}^* , ~~~ m_D^T M_D^{-1} \propto   
\mathbbm{1}, ~~~~U_S = U_{13}^{max}. 
$$
So,  $U_{\omega}$ appears from 
the charged leptons mass matrix 
whereas the hidden sector gives maximal 1-3 mixing, 
and there is a complete screening. 
As a result,  $U_0 \propto U_\omega U_{13}^{max} = U_{TBM}$. 

\begin{figure}[t]
\centering\includegraphics[width=4.0in]{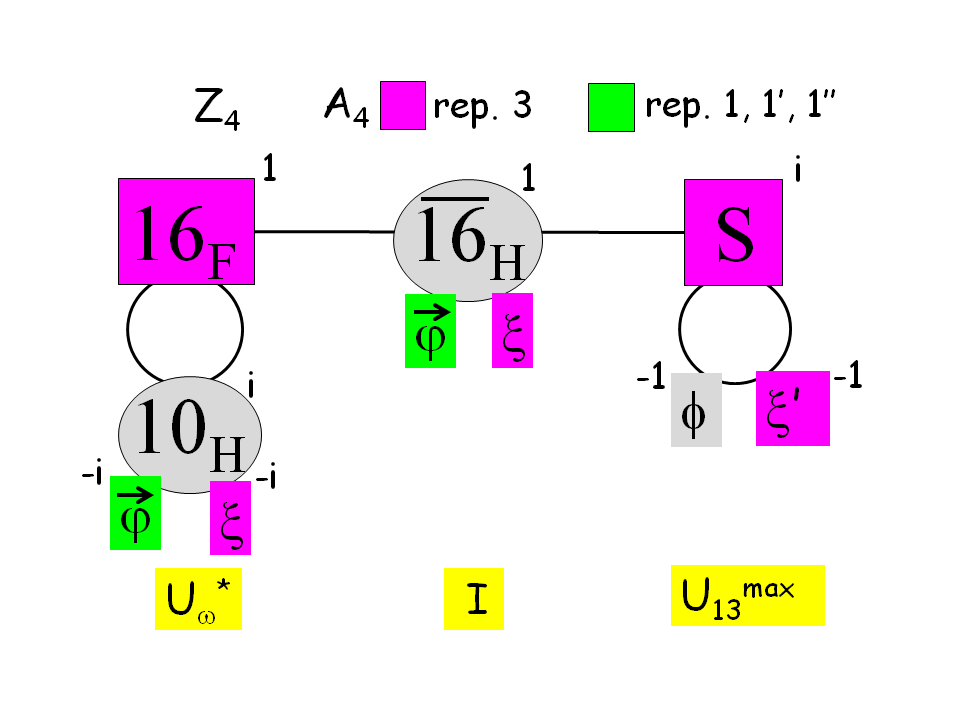}
\vskip -1cm
\caption{The same as in Fig.~\ref{fig:ss-scheme} but for the mixed flavon scheme for 
the TBM mixing.}
\label{fig:mix1-scheme}
\end{figure}

The symmetry  assignments of the field content we use is
$$
\{16_F,\, S \,|\, 10_H, \, \overline{16}_H \,|\, \vec{\varphi}, \,\phi, \,\xi, \,\xi' \} \sim  \{1, \, i\, |\, i, \,1 | - i, - 1, -i, -1 \}
$$
under $Z_4$, and the full graphic representation are shown in Fig.~\ref{fig:mix1-scheme}. 

Both the singlet ${\vec \varphi}$  and triplet $\xi$ flavons, with arbitrary VEVs, participate  
in visible and portal interactions. The others, $\phi$ and $\xi'$,  exclusively couple with fermion singlets.

The Lagrangian has the following terms 
\begin{eqnarray}
{\mathcal L_{D}} &=&  \frac{10_H}{\Lambda_f}  \left[\vec{y}_{10}\, (16_F \cdot 16_F)_\varphi  
\vec{\varphi} + 
h_{10} \,(16_F \cdot 16_F  \cdot \xi) \right] ,   \,  
\nonumber\\
{\mathcal L_{NS}} &=&    \frac{\overline{16}_H}{\Lambda_f} 
\left[ \vec{y}_{16}\, (16_F \cdot S)_\varphi \vec{\varphi} +   
h_{16} \,(16_F \cdot S \cdot \xi)  \right]\, , 
\nonumber\\
{\mathcal L_{S}} &=& y_{1}\, (S \cdot S) \, {\phi} +  
h_{1}' \,(S \cdot S \cdot \xi') \, . 
\end{eqnarray}
The  $Z_4$ symmetry forbids all the interactions without flavons including   
the bare mass terms.\\ 

For the Dirac mass matrix we obtain  
$$
m_D =  {\langle 10_H \rangle \over \Lambda_f}
\begin{pmatrix}
M_{1}  &  h_{10} u_{3} &  h_{10} u_{2}\\
...   & M_{2}  &  h_{10} u_{1}\\
...  & ...  & M_{3}
\end{pmatrix}, 
$$
where $M_{i}$ are defined in Eq.~(\ref{Mi-def}).  
To be diagonalized by $U_\omega$ this matrix  should be of the form 
Eq.~\eqref{omega:ml}. 
This requires that $M_i = h_{10} u_{i}$, or 
according to Eq.~\eqref{Mi-def},  
the relation (\ref{uv-correlation}) with  $h = h_{10}$ and $\vec y = \vec y_{10}$.

Using this VEV relation, the  mass matrix can be written as 
$$
m_D =  {h_{10} \langle 10_H \rangle \over \Lambda_f}
\begin{pmatrix}
u_{1}  &  u_{3} &   u_{2}\\
...   & u_{2}  &   u_{1}\\
...  & ...  & u_{3}
\end{pmatrix}.
$$
It is diagonalized as $U_l^\dagger m_D V_R$, where $U_l = U_\omega^*$ 
and $V_R = U_\omega$.
The eigenvalues of the matrix equal
$$
(m_{D1}, m_{D2}, m_{D3})^T =  {h_{10} \langle 10_H \rangle \over \Lambda_f} 
(\sqrt{3} U_\omega^*) (u_{1}, u_{2}, u_{3})^T =  
{3 \langle 10_H \rangle \over \Lambda_f} (y_{10} v,~ y_{10}' v', ~ y_{10}'' v'')^T ,  
$$
so the mass hierarchy requires    that $y_{10}v \ll y'_{10}v' \ll y''_{10} v''$.\\

The same flavons $\vec \varphi$  and $\xi$ appear 
in ${\mathcal L_{D}}$ and ${\mathcal L_{NS}}$.
Therefore, the flavor structures of mass matrices $m_D$ and $M_D$ 
are the same provided that the Yukawa couplings of $10_H$  and $16_{H}$ are proportional to
each other 
\begin{equation}
{y_{10} } : {y'_{10} } : {y''_{10} } : { h_{10} } = 
{y_{16} } : {y'_{16} } : {y''_{16} } : { h_{16}}.
\label{eq:prop2}
\end{equation}
That is, now couplings of $\vec \varphi$  and $\xi$ should also correlate.  
Possible origins of this proportionality have been discussed in sect.~\ref{sect:ssss}.
As a consequence,  
\be
m_D^{\nu} = m_l =  m_D 
= {y_{10}\langle 10_H \rangle \over y_{16}\langle \overline{16}_H \rangle} M_D,  
\label{eq:unitportal}
\ee
and therefore  ${m_D M_D^{-1\,T}} =  
y_{10}\langle 10_H \rangle / (y_{16}\langle \overline{16}_H \rangle) \cdot {\mathbbm 1}$,  
reproducing the complete screening.\\

Maximal 1-3 mixing should be obtained from  $M_S$ in the form of $M_{33}$. 
If $\langle{\phi}\rangle =  v_1$ and 
the VEV of the triplet is aligned as  
$\langle{\xi'}\rangle \to (0, u', 0)$, we obtain 
$$
M_S =
\begin{pmatrix}
y_1 v_1    & 0 &  h_1' u'\\
...   &y_1 v_1    &  0\\
... & ...  & y_1 v_1
\end{pmatrix}.   
$$
The eigenvalues of $M_S$ equal  $y_1 v_1 \pm h' u'$ and $y_1 v_1$, and consequently,  
the light neutrino masses equal 
\be
m_{1,3} =  B (y_1 v_1 \pm h_1' u'), ~~~~
m_2     =  B y_1 v_1, ~~~~~~~~
B \equiv { y^2_{10}\langle 10_H \rangle^2 \over y^2_{16}\langle \overline{16}_H \rangle^2}.   
\label{modelB:nmasses}
\ee
This gives normal mass ordering 
with $m_1 : m_2 : m_3 \simeq  1 : 1 : 3$, 
if  $ h_1' u' \sim -2 y_1 v_1$.   
The inverted mass ordering  can be obtained if 
two additional  flavon singlets, $\phi'$ and $\phi''$,  are introduced 
in the hidden sector with  couplings satisfying
$y_1'\langle \phi' \rangle =\omega  y_1''\langle \phi'' \rangle $. This adds 
one free parameter  to $M_S$, and thus,  allows arbitrary neutrino masses 
without changing maximal 1-3 mixing. \\

The model can be simplified if certain high-order corrections to $m_D$ are added.
Assume that  $\langle \varphi' \rangle = \langle \varphi'' \rangle  = 0$,
or simply that  $\varphi'$ and $\varphi''$ are removed from the scheme. 
In practice, this is equivalent to the case $\vec{y}_{10} = 0$ but $m_D^0 \neq 0$. 
If then  
$$
\langle{\xi} \rangle =  u (1,~ 1,~ 1), 
$$
we obtain 
$$
m_D = M_{33}^{special} 
$$
of Eq.~(\ref{special}) with $\mu = y_{10} v $ and $\beta = h_{10} u$. 
The equality ~\eqref{eq:unitportal}  and  $M_S$ remain unchanged.  

To break the degeneracy of eigenvalues of $m_{D}$ and fix  $U_l = U_\omega$,
one has to introduce a non-symmetric correction $\delta m_D$
(see e.g. $\Delta M'$ in Eq.~\eqref{perturbms}). In particular, one can 
use the anti-symmetric correction 
$$
\delta m_D = \delta m
\begin{pmatrix}
0    & 1  &  -1\\
-1   & 0 & 1 \\
1  & -1   &  0
\end{pmatrix} 
$$
which  originates from  the effective $120_H$. 
The latter appears  from non-renormalizable interactions 
$$
\frac{1}{\Lambda} 16_F 16_F 16_H 16_H, ~~~~~ 16_F 16_F 45_H 10_H,  
$$
noticing that the decomposition of ($16_H\times 16_H$) 
contains $120_H$, 
see \cite{Babu:1998wi, Altarelli:2002hx, Morisi:2007ft} for examples. 
As a result, $U_{PMNS} = U_{\omega} U^\text{max}_{13}$ 
with non-degenerate masses of charged leptons given by
$$
m_{D1} = y_{10}  v  + 2 h_{10}  u, ~~~
m_{D2,D3}= y_{10} v  -  h_{10}  u \mp i \sqrt{3} \delta m.
$$
To  keep screening, similar corrections should also be introduced for $M_D$. 
It is straightforward to see that the masses
of light neutrinos are the same as in Eq.~(\ref{modelB:nmasses}),
because $M_S$ is unchanged.

\subsection{Scheme with mixed contributions: the BM mixing}

Our framework also allows for new possibilities to produce the BM mixing (see Appendix.~\ref{App:mixings}).
The BM mixing can be factorized by two maximal 
mixing rotations: $U_{BM} = U_{23}^{max} U_{12}^{max}$. 
In this connection we present a scenario  
in which  maximal 2-3 rotation comes
from the charged leptons and the maximal 1-2 one arises from the hidden sector: 
$$ 
U_l = (U_{23}^{max})^T, ~~~~~ m_D (M_D)^{-1} \propto {\mathbbm 1}, ~~~~~ U_S = U_{12}^{max}. 
$$

\begin{figure}[t]
\centering\includegraphics[width=4.0in]{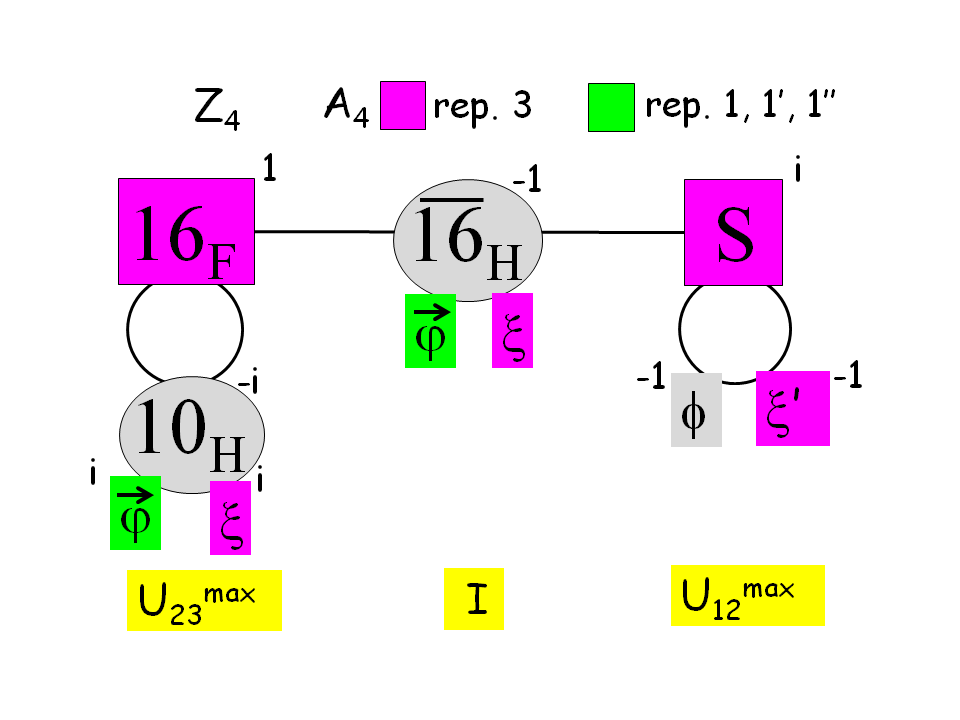}
\vskip -1cm
\caption{The same as in Fig.~\ref{fig:ss-scheme} 
but for  the mixed flavon scheme for the BM mixing.}
\label{fig:mix2-scheme}
\end{figure}

The particle content and the charge assignments are given
in Fig.~\ref{fig:mix2-scheme}, where the $Z_4$ charges are 
$$
\{16_F, \,S \,|\, 10_H, \, \overline{16}_H \,| \,\vec{\varphi},\, \phi,\, \xi, \,\xi' \} \sim  \{1,  i\,| - i, - 1 |\, i,-1,\,i, -1 \},
$$ 
respectively. All the mass matrices are of the type $M_{33}$.  
Bare mass terms and operators without flavons  
are forbidden by $Z_4$ symmetry.
The symmetry assignment  allows flavon singlets 
$\vec{\varphi}$ and triplet $\xi$ 
to couple with fermions in  visible and portal interactions,
while the other two flavons, singlet $\phi$ and triplet $\xi'$, appear in the hidden sector only.

To  get maximal 2-3 mixing from the charged leptons, where $m_l = m_D$, 
we impose equalities
\begin{equation}
y'_{10}  =  y''_{10}, ~~~~~  v'  =  v'',~~~~~ \langle \xi \rangle =  (u , 0, 0).
\label{BMpattern:assumption2}
\end{equation}
These equalities  imply the permutation symmetry (\,$' \leftrightarrow \,''$\,) in the visible and portal sectors, 
and therefore lead to  maximal 2-3 mixing.
Under conditions (\ref{BMpattern:assumption2}) we obtain
$$
m_D = 
{\langle 10_H \rangle \over \Lambda_f}
\begin{pmatrix}
y_{10}v + 2y_{10}' v'    & 0  &  0\\
0   & y_{10}v - y_{10}' v'  & h_{10} u\\
0      & h_{10} u  &  y_{10}v - y_{10}' v'
\end{pmatrix} . ~~~
$$
The eigenvalues of this matrix equal
\begin{equation}
m_{D1} = {\langle 10_H \rangle \over \Lambda_f} (y_{10}v + 2y_{10}' v'),~m_{D2,\,D3}
= {\langle 10_H \rangle \over \Lambda_f}( y_{10}v - y_{10}' v' \mp h_{10} u).
\end{equation}
To obtain  mass hierarchy we should arrange that
$y_{10}' v' \approx - 0.5 y_{10}v$,  which leads to
$$
m_{D2} \simeq {\langle 10_H \rangle \over \Lambda_f}  
\left(\frac{3}{2}y_{10}v   - h_{10} u \right), ~~~~~
m_{D3} \simeq {\langle 10_H \rangle \over \Lambda_f} 
\left( \frac{3}{2} y_{10}v + h_{10} u \right),
$$
and $m_{D1} \ll m_{D2}$. Additional  relation
$h_{10} u \approx \frac{3}{2}y_{10}v $  gives  $m_{D2} \ll m_{D3}$. 

To realize complete screening, $m_D \propto  M_D$, we  
impose the proportionality of both singlet and triplet couplings $\vec y$ and $h$
with $10_H$-plets and $\overline{16}_H$-plets as Eq.~(\ref{eq:prop2}).  

The maximal 1-2 mixing from $M_S$ requires that the second flavon triplet, 
interacting in the hidden sector, has VEV $\langle{\xi'} \rangle\to (0, 0, u')$,
which implies  the permutation symmetry of the first and second  generations in hidden sector.
Equal diagonal elements of $M_S$ are generated by 
singlet $\phi$ with  VEV $\langle \phi \rangle =  v_1$.
Under these conditions 
$$
M_S  =
\begin{pmatrix}
y_1 v_1   & h_1' u' &  0\\
 h_1' u'    & y_1 v_1    & 0\\
0 & 0  &  y_1 v_1
\end{pmatrix}.  ~~~
$$
Since $m_\nu \propto M_S$, an
agreement with observed mass spectrum can be achieved for the inverted 
mass hierarchy  when $|y_1 v_1|  \ll |h_1' u'|  $. As a result, 
the three light neutrino masses equal   
$m_3 \propto y_1 v_1 $,
$m_2 \approx m_1 \propto  h' u'$ and $\Delta m^2_{21} \propto 4 y_1 v_1  h_1' u'$.

To obtain normal mass hierarchy we can  introduce also $\phi'$ and $\phi''$
with couplings $y_1'\langle \phi' \rangle =\omega  y_1''\langle \phi'' \rangle $,
which actually allows to  obtain arbitrary values of neutrino  masses. 
Recall that conditions we impose here can be consequences of additional symmetries 
operating in the hidden sector. 

\section{Schemes with $S$ transforming as singlets of $A_4$ } \label{sect:sing}

For $\vec{S} = (S, S', S'') \sim ({\bf 1},\, {\bf 1'},\,{\bf 1''})$ 
the portal and the hidden parts of the Lagrangian 
change with respect to Eq.~(\ref{s-triplet}): 
\begin{eqnarray}
{\mathcal L_{NS}} &=&   
\frac{\overline{16}_{H}}{\Lambda_f} \, \vec{h}_{16} 
\,  (16_F \cdot \xi)_S  \,\vec{S},  
\nonumber\\
{\mathcal L_{S}} &=&  M^0_{S} S S  
+ 2 \tilde{M}_S^0 \, S' S'' +  
( y_{11}\,S S + 2 y_{23} \, S'S'')  \, \varphi \notag\\
&+&  (y_{22}\, {S'}  {S'} + 2 y_{13}\, {S} { S''}) \, 
\, {\varphi'}+  ( 2 y_{12}\, { S} {S'} + y_{33}\, {S''} {S''} ) \, {\varphi''}.
\end{eqnarray}
The visible interactions are the same as before, so that 
the Dirac mass matrices have two contributions: one from singlets  $\vec{\varphi}$ and  
the other from triplet $\xi$ of $A_4$.  
Since  the fermion singlets $S$ transform as the three one-dimensional 
representations of $A_4$, the portal term has only one contribution: 
from triplet $\xi$. 
Therefore, in contrast to the schemes in the previous section with 
$S\sim {\bf 3}$, now we can only have two possibilities for zero couplings related 
to the Dirac terms: 
${h}_{10}  = 0$ or $\vec{y}_{10}  = 0$.  For both possibilities, according to   
Eqs.~(\ref{eq:m31}-\ref{mass:form11})  the portal and hidden sector 
mass matrices are  given by  (in the case of only one  triplet flavon)  
\be 
\label{m31:poss1}
M_D = M_{31} = \sqrt{3} D_u U_\omega D_h,~~~~~~~M_S = M_{11}.   
\ee

Since $16_F$  and $S$ are in different 
representations of $A_4$, 
the corresponding mass matrices have different flavor structures and 
it becomes non-trivial to get certain correlations between them.   

\subsection{Scheme with ${h}_{10}  = 0$}

Vanishing ${h}_{10}$ can be achieved, e.g.  due to $Z_2$ symmetry, 
under which only $\xi$ and $\overline{16}_H$ are odd.  
Only singlet flavons contribute to the visible sector masses 
and therefore  $m_D =  m_D^0 {\mathbbm 1} + M_{33}^{\varphi} $    
with elements fixed by masses of charged leptons (or up-type quarks), 
{\it i.e.} $m_D = m_l =  m_l^{diag}$.  
So, mixing is generated by neutrinos. 
The light neutrino mass matrix can be written as  
$$
m_\nu =  m^{diag}_l (M_{31}^T)^{-1} M_S ~(M_{31})^{-1} m^{diag}_l,  
$$
or explicitly, according to Eq.~(\ref{m31:poss1})
\be
m_\nu =  \frac{1}{3} {m^{diag}_l \over D_u} U_\omega^* {1\over D_h} M_S 
{1\over D_h} U_\omega^* {m^{diag}_l \over D_u}. 
\label{nunu3}
\ee

For  $D_h \propto {\mathbbm 1}$ and 
$m^{diag}_l (D_u)^{-1} \propto {\mathbbm 1}$   
(partial screening),  the expression (\ref{nunu3}) is reduced to  
$$
m_\nu \propto  \frac{1}{3} U_\omega^* M_S U_\omega^* . 
$$
If the bare mass term dominates in $M_S$ we would get 
$$  
m_\nu = \frac{1}{9} \left[ 3\tilde{M}_S^0 {\mathbbm 1} + 
\left({M}_S^0 - \tilde{M}_S^0 \right) M_{dem}\right],  
$$
where  $M_{dem}$ is the matrix with all the elements equal 1. 
It can be diagonalized by  the TBM mixing but has two equal neutrino masses. 
Such a degeneracy means that the mixing is not fixed completely. 
More contributions are required to break it. 
A straightforward possibility is to use additional contributions 
to $M_{11}$ in order to obtain 
maximal 1-3 mixing from $M_S$, so that $U_0 = U_\omega U_{13}^{max} = U_{TBM}$. \\

In what follows we will consider another way of getting the required 
corrections, which also justifies some assumptions we  have
made before. We elaborate on a scenario 
which realizes relations 
$$
m_D = diag,~~~m_D^T M_D^{-1} \propto U_{\omega}^*,~~~D_h^{-1} M_S D_h^{-1} = M_{13}^{max}.
$$
Here, as in the previous case,  the first relation means that $U_l = {\mathbbm 1}$, 
the second relation  corresponds to  
partial screening of the Dirac structures  
which generates the magic matrix. 
The singlet matrix together with $D_h$ produces 
maximal 1-3 mixing.  
Then 
$U_0 \propto U_{\omega}\,U_{13}^{max} = U_{TBM}$.

\begin{figure}[t]
\centering\includegraphics[width=4.0in]{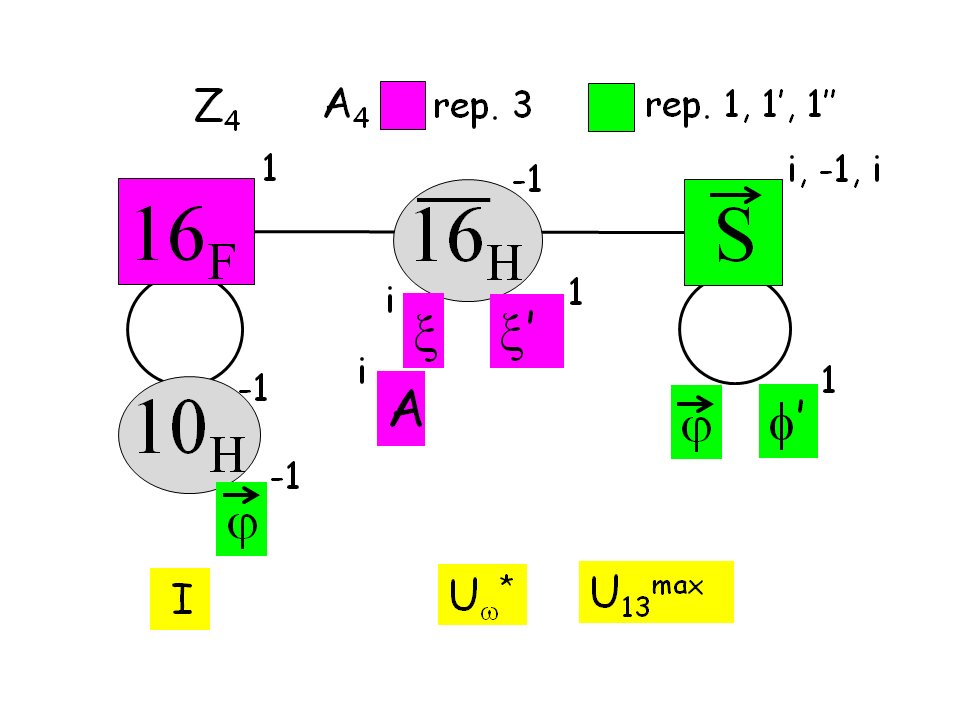}
\vskip -1cm
\caption{The same as in Fig.~\ref{fig:ss-scheme}
but for the flavon scheme  with $S$ in three singlet  
representations of $A_4$  and $h_{10}  = 0$.
}
\label{fig:g10-scheme}
\end{figure}

We introduce $Z_4$ to forbid $h_{10}$ and  some other  couplings.   
The field content,  symmetry assignments and 
graphic representation of the scheme are given  in Fig.~\ref{fig:g10-scheme}. 
The features of the scheme 
(as compared with minimal structure discussed in the beginning of the subsection)   
are (i) different $Z_4$ charges of fermion singlets, (ii) two flavon triplets  
participating in the portal interactions, 
(iii) additional flavon singlet  $\phi'$, (iv) auxiliary flavon triplet $A$.

The most general expressions for the three  terms of the Lagrangian
(\ref{eq:larg}) allowed by symmetry in the lowest order 
are
\begin{eqnarray}
{\mathcal L_{D}} & = &
\frac{10_H}{\Lambda_f} \vec{y}_{10} (16_F \cdot 16_F)_\varphi \vec{\varphi},  
\nonumber\\
{\mathcal L_{NS}} & = &
\frac{\overline{16}_H}{\Lambda_f}
\left[h_{16}\, (16_F \cdot \xi) \, S +
h'_{16} \,(16_F \cdot \xi^\prime)'' \, {S'} +
h''_{16} \, (16_F \cdot \xi)' \,{S''} \,\right] ,
\nonumber\\
{\mathcal L_{S}} &=& y_{1}\, (S \cdot S) \, \varphi 
+ 2 y_{1}' \,({S} \cdot {S''} ) \,  {\varphi'}\, 
+  y''_{1}\, (S'' \cdot S'') \,{\varphi''}  
+ \tilde{y}'_{1}\, (S' \cdot S') \, {\phi'}.
\label{eq:hidden}
\end{eqnarray}
Notice that  $Z_4$ forbids the low-dimension flavonless  
term $16_F 16_F 10_H$ since $10_H$ has the charge $-1$. 
It also ensures that triplet flavons do not couple with
$16_F 16_F$, and therefore only singlets contribute to the
Dirac mass matrix, leading to $m_D = M_{33}^{\varphi}$.
The symmetry $Z_4$ also forbids the bare mass terms $(S S)$ and $(S' S'')$.

There is a generic problem here: On the one hand, 
obtaining $U_\omega$ from  $M_D$  
requires (in the simplest case) 
that only one flavon triplet participates in the portal  
interactions.  
On the other hand, to  get maximal 1-3 mixing from the 
hidden sector one needs to distinguish $S'$ from $S$ and $S''$,  
so that the terms $(S S')\varphi''$
and $(S' S'')\varphi$ are forbidden. 
To forbid such terms, singlets  $S$ and $S''$ should have equal $Z_4$ charges which differ 
from the charge of $S'$. Therefore in ${\mathcal L}_{NS}$ two flavon triplets
$\xi$ and  $\xi'$ carrying different  $Z_4$  charges
are necessary. 
Moreover, a new flavon singlet, $\phi^\prime$, with 
$Z_4$ charge different  from $\vec{\varphi}$, 
should be introduced  to generate the  mass of $S'$. \\

Now in the visible sector, both the bare mass and off-diagonal elements of $M_{33}$ are absent,
so ${\mathcal L_{D}}$ in Eq.~(\ref{eq:hidden}) only generates
the same diagonal Dirac mass matrices $m_D = M_{33}^{\varphi} (\vec{y}_{10}, \vec{v})$
to all SM fermion components as in Eq.~(\ref{Mi-def})
\begin{eqnarray}
(m_{D1},~ m_{D2}, ~ m_{D3})^T =  \frac{\langle 10_H \rangle}{\Lambda_f} 
\sqrt{3} U_\omega (y_{10} v,~  y_{10}' v', ~ y_{10}'' v'')^T.
\end{eqnarray}
So the expressions of mass matrices are in the flavor basis from the beginning 
\footnote{The diagonal matrix $m_D$ can also be written
in the form of product of  matrices as
$\tfrac{\langle 10_H \rangle}{\Lambda_f}
U_{\omega}^* (y_{10} v {\mathbbm 1} + y_{10}' v' C + y_{10}'' v'' C^T )
U_{\omega}$, where $C$ is the cyclic matrix with only 
non-zero (1,3), (2,1), and (3,2) unit elements.}. The hierarchy of mass 
eigenvalues $ m_{D1}, m_{D2}, m_{D3}$ can be obtained in the same way as Eq.~\eqref{eq:condhier}.\\
 
The portal mass matrix generated by the terms ${\mathcal L_{NS}}$ equals 
\begin{equation}
M_D =
\frac{\langle \overline{16}_H \rangle}{\Lambda_f}
\begin{pmatrix}
h_{16} u_1  & h_{16}'  u_1'   &  h_{16}'' u_1 \\
h_{16} u_2  & h_{16}'  u_2'\omega   &  h_{16}'' u_2 \omega^2 \\
h_{16} u_3  & h_{16}' u_3'\omega^2   &  h_{16}'' u_3 \omega
\end{pmatrix},
\label{eq:martixMD}
\end{equation}
where the VEVs of two flavon triplet are denoted as   
$\langle{\xi}\rangle =  (u_1,u_2, u_3)$, 
$\langle{\xi^{\prime}}\rangle  =  (u_1', u_2', u_3')$.
To get exactly the form of $M_{31}$, their VEVs need to be aligned 
in the same way:  
$\langle{\xi}\rangle   =  \langle{\xi'}\rangle \,$
\,({\it i.e.} $u_i = u_i'$). With this equality, 
the mass matrix of Eq.~(\ref{eq:martixMD}) can be written as 
in Eq.~\eqref{eq:m31}: $M_D = 
{\sqrt{3}} {\langle \overline{16}_H \rangle} D_u U_{\omega} D_{h} /{\Lambda_f}$. 

The mass matrix of singlet fermions generated by
Eq.~(\ref{eq:hidden}) has the form of $M_{11}$  with zeros due to symmetry assignment:
\begin{equation}
M_S =
\begin{pmatrix}
y_{1} v   & 0    &  y_1' v'  \\
...  & \tilde{y}'_{1} v_1   &  0 \\
... &  ...  &  y_{1}'' v'' 
\end{pmatrix}, 
\label{eq:martixMS}
\end{equation}
where $v_1 \equiv \langle{\phi'}\rangle $. 

Combining $M_D$ and $M_S$  
we obtain from   Eq.~\eqref{eq:matrix3-1} the light neutrino mass matrix 
\be 
\label{modelA:Mnu}
m_\nu = {1\over 3}\,\left(\frac{\Lambda_f}{\langle \overline{16}_H \rangle} \right)^2 {m_D^{diag} \over D_u} 
\cdot U_\omega^* \cdot
\begin{pmatrix}
\frac{y_{1} v}{h_{16}^2}    & 0    &  \frac{{y}_1' v' }{h_{16} h_{16}''}   \\
...  & \frac{ \tilde{y}'_{1} v_1}{h_{16}'^2}   &  0 \\
...     &  ...   &  \frac{y_{1}'' v''}{h_{16}''^2}
\end{pmatrix}
\cdot U_\omega^* \cdot {m_D^{diag} \over D_u}.
\ee

To get the TBM mixing the following two conditions should be fulfilled:
\be
m_D^{diag}  \propto  D_u,
\label{eq:fcond}
\ee
which leads to partial screening,   and
\be
\frac{y_{1}}{h_{16}^2} v = \frac{y_{1}''}{h_{16}''^2} v'',
\label{eq:second}
\ee
to get maximal 1-3 mixing from the central matrix  of Eq.~\eqref{modelA:Mnu}.

The first condition (\ref{eq:fcond})  
is the correlation of the singlet and triplet VEVs, 
which can be obtained in the same way as it was done in 
Eq.~(\ref{aux:cancel}), by introducing the auxiliary $A_4$ triplet field 
$A$ with VEVs  $\langle A \rangle = (a,\,a,\,a)/\sqrt{3}$. 
The inequality $m_D \ll M_D$ implies a very small VEV:  $a \ll 1$.  
Therefore the auxiliary field does not change other structures of the model.
Provided that $y_{10} = y'_{10}= y''_{10}$, the Dirac masses can be rewritten as 
$$
(m_{D1},  m_{D2}, m_{D3}) = 
  \frac{\sqrt{3} a \,y_{10} \langle 10_H \rangle}{  \Lambda_f} (u_1, u_2, u_3). 
$$

The second condition, Eq.~(\ref{eq:second})  can be satisfied if, e.g. 
$$
h_{16}  = \omega h''_{16},\, \, \, \, \, \, y_{1}  
= \omega y''_{1}, \, \, \, \, \, v = \omega v''.
$$
These equalities then produce the masses of light neutrinos as
$$
m_{1, 3}  = C \frac{1}{h_{16}} \left(\frac{y_{1}}{h_{16}}
\pm  \frac{y'_1}{h_{16}''} \right) v,
~~~~ m_2 = 
C \frac{ \tilde{y}'_{1}}{h_{16}'^2} v_1, ~~~~~
C \equiv \frac{a^2  \langle 10_H \rangle^2 y_{10}^2}{ \langle \overline{16}_H \rangle^2}.  
$$
Strong mass hierarchy requires $y_1/h_{16} \approx  \pm {y}_1' /  h_{16}''$.\\

There exist different ways to avoid the introduction of the second triplet and tuning 
of its VEVs. One possibility is to use single $\xi$ with $Z_4$ charge $i$
and assign to $S'$ the same charge $i$ as other singlets have.  
Then to get nearly maximal 1-3 mixing  from the hidden sector 
additional  symmetry, e.g. $Z_2$, can be introduced and made conserved 
in the hidden sector only (see also discussion in sect.~\ref{TBM:Factorizing}). If 
$S'$ is odd and other  particles of the hidden sector are even, 
then $M_S$ will have the form (\ref{eq:martixMS}) 
with zero 12- and 23- elements.   
This additional symmetry is broken in the visible and portal  sectors, 
which have substantially lower energy scale $M_{GUT}$. 
So, one expects that corrections to the $Z_2$ symmetric results are of the order 
$M_{GUT}/ M_{Pl}$, which are not important.

Instead of second flavon triplet $\xi'$,  
second $16'_{H}$ with $Z_4$ charge $i$ 
can be introduced in the portal interaction  
$$
\tfrac{h'_{16}}{\Lambda_f} (16_F \xi)'' 16'_{H} S'. 
$$
The  VEVs of 16-plets should be equal, $\langle 16_{H}' \rangle =  \langle 16_{H} \rangle$, 
to get the matrix factor $U_\omega$.  Alternatively,  to replace $\xi'$, one can also introduce  
a flavon singlet $\phi$ with  $Z_4$ charge $-i$, which leads to a 
dimension-6 portal interaction $(16_F \xi)'' 16_{H} \phi S'$. 

Notice that this scheme can be made more logical if  three $\overline{16}_{H}$ (or three $\xi$) with 
different $Z_4$ charges are introduced.

\subsection{Scheme with $\vec{y}_{10}  = 0$ }

\begin{figure}[t]
\centering\includegraphics[width=4.0in]{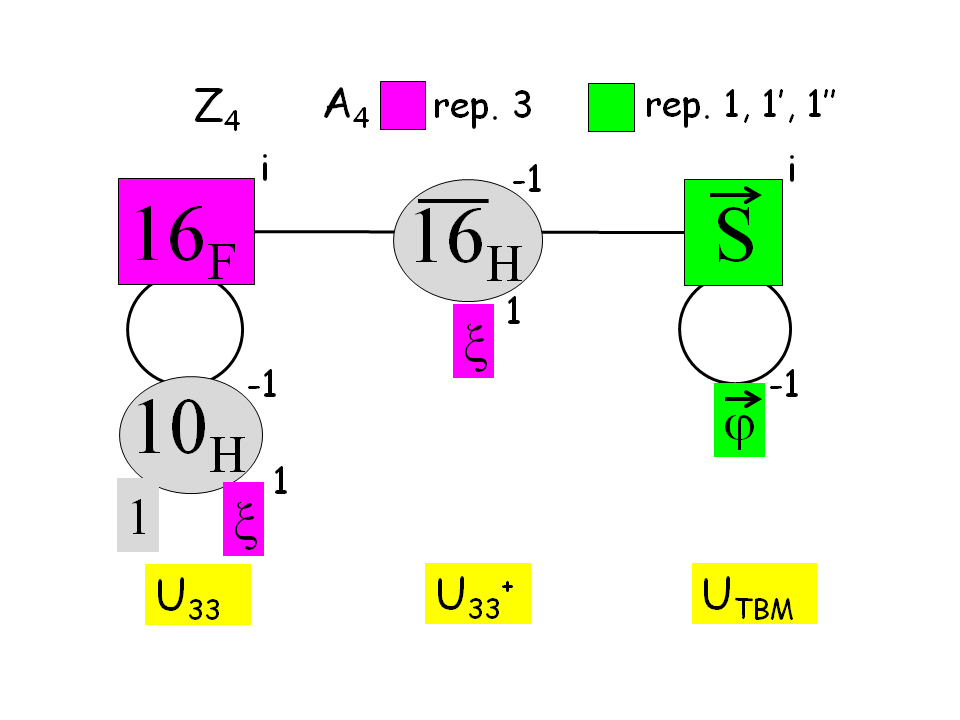}
\vskip -1cm
\caption{The same as in Fig.~\ref{fig:ss-scheme} 
but for the flavon scheme  with $S$ in three singlet 
representations of $A_4$  and $\vec{y}_{10}  = 0$.
}  
\label{fig:h10-scheme}
\end{figure}

The  visible part of the Lagrangian is  as in Eq.~(\ref{s-triplet}) 
with $\vec{y}_{10}  = 0$ and $h_{10} \neq 0$. The particle content and symmetry 
alignment are shown in Fig.~\ref{fig:h10-scheme}.   
The triplet flavon contributes to both visible and portal interactions, but 
structures of $m_D$ and $M_D$ are different. 
Vanishing  $\vec{y}_{10}$ can be obtained with $Z_4$ assignment   
$$ 
\{16_F,~S |10_H,~\overline{16}_H | \vec{\varphi},~  \xi \} 
\sim \{ i,~ i | -1, ~ -1 |  - 1, ~ 1 \}, 
$$
as in the model of sect.~\ref{sect:ssss}. Flavor invariant mass $m_D^0$ is allowed. 
Consequently,  $m_D = m_D^0 {\mathbbm 1}  + M_{33}^{\xi}$,  
as in   Eq.~(\ref{mass:trip})  
and  $m_l = m_D$. It is thus diagonalized by $U_{33}$.  
The portal mass matrix $M_D$ has the form of $M_{31}$. 

The neutrino mass matrix equals 
\be
m_\nu =  \frac{1}{3} m_D {1\over D_u} U_\omega^* {1\over D_h} M_S
{1\over D_h}  U_\omega^* {1\over D_u} m_D , 
\label{nunu4}
\ee
and it converts to 
\be
m_\nu =  \frac{1}{3} m_D^{diag} \cdot  U_{33} {1\over D_u} U_\omega^* \cdot {1\over D_h}   M_S 
{1\over D_h} \cdot U_\omega^* {1\over D_u} U_{33} \cdot  m_D^{diag},
\label{nunu5}
\ee
in the flavor basis. In the case of  two Higgs 10-plets, there is actually $m_D^{diag} = m_u$.  
In the hidden sector the bare mass terms, $M_S^0$ and  $\tilde{M}_S^0$, are forbidden.

If  $D_u = {\mathbbm 1}$,  the matrix  $M_{33}$ has  special form 
$M_{33}^{special}$, which can be diagonalized by 
$U_{33} = U_{\omega}$. In this case Eq.~(\ref{nunu5}) gives 
\be
m_\nu \propto   \frac{1}{3} {m_D^{diag}\over D_h} M_S  {m_D^{diag}\over D_h} .
\label{nunu7}
\ee
Then for partial screening, $m_D^{diag} (D_h)^{-1} \propto {\mathbbm 1}$,  we obtain 
$m_\nu \propto M_S$.   With these conditions we essentially 
reproduce complete screening. Then $M_S$ should have the TBM form, which can be obtained 
in the way described by Eqs.~(\ref{mass:form11}, \ref{tbm11}) in sect.~\ref{sect:IID}. 

\section{Generalizations} \label{sect:general}

We have assumed above that the light neutrino masses 
are mainly determined by the double seesaw mechanism with three fermion singlets $S$.  
Here we consider other possible contributions to masses and mixing, 
including the CKM physics effects. 

\subsection{Effects of linear seesaw} 
\label{chapt:gene}

The direct mixing between left-handed neutrinos and fermion 
singlets $S$ via the $\overline{16}_H$ Higgs gives the linear seesaw 
contribution to the light neutrino masses:
$$
m^{ls}_\nu = - m_D {1\over M_D^T} m'^T -  m' {1\over M_D}  m_D^T,
$$
which are the last terms of Eq.~\eqref{neutrino:2mass}. Under the condition  $M_S \gg M_\text{cr}$,  
where $M_\text{cr}$ is defined by Eq.~\eqref{eq:crit},  
this contribution is  subleading,  but  may provide certain required 
corrections to some observables.

On the contrary, if  $M_S \ll M_\text{cr}$, the linear seesaw contribution 
is dominant. As we have mentioned in sect.~\ref{sect:IIB} in the presence of only 
one $\overline{16}_H$, $m' \propto M_D$,  and therefore $m^{ls}_\nu  \propto m_D$.  
Thus,  no mixing is generated. If more $\overline{16}_H$ 
multiplets are introduced, the proportionality can be broken: 
One $\overline{16}_H$ with its VEV along the $SU(2)_L$ singlet direction, $\langle \overline{16}_H \rangle_1 $,  
will be responsible for  $M_D$ and another one, $\overline{16}_H'$, with its  
VEV along the  $SU(2)_L$ doublet direction, $\langle \overline{16}_H \rangle_d$, 
generates the mass $m'$. If couplings of two $16$-plets are different, 
$y_{16} \neq y_{16}'$, the matrices $m'$ and $M_D$  have different structures, and  
non-vanishing neutrino  mixing can be generated. 
Notice that for the linear seesaw  $M_S$ is irrelevant and  $m'$ plays the role of $M_S$.

As an example, we will briefly describe the case when the linear seesaw gives 
the main contribution to neutrino mass. 
We assume that $16_F$ and $S$ are $A_4$ triplets. Then  $m_l = m_D$ can  have 
the form of $M_\omega$, as Eq.~\eqref{omega:ml}, which is diagonalized by $U_\omega$. 
Also complete  screening can be realized,  
$m_D  M_D^{-1\,T} \propto {\mathbbm 1} $.  For $m'$ we can have  
$$
m' = M_0  {\mathbbm 1}  + M^{\xi}_{33}. 
$$
For certain VEV alignment as in the scheme  
of sect.~\ref{TBM:33mixing}, $m'$ has the form of Eq.\eqref{eq:unu1}, 
producing the  maximal 1-3 rotation, so that 
$U_0 = U_\omega U_{13}^{max} = U_{TBM}$.

Realization of partial screening is different from the case of  double seesaw. 
Let us take a scheme with $16_F \sim {\bf 3}$ 
and $\vec S \sim ({\bf 1}, {\bf 1'}, {\bf 1''})$, 
in which  $m_D$ is  diagonal,  $m_D  M_D^{-1\,T} \propto U_\omega^* $, and   
$m'$ has the form of $M_{31}$. Then  the  
neutrino mass matrix is given by  
$$
m^{ls}_\nu \propto U_\omega^* m'^T +  m' U_\omega^* \propto U_\omega^* D_h U_\omega D_u + 
D_u U_\omega D_h U_\omega^* . 
$$
Under conditions  $u_3= u_2$ in $D_u$ and $h_3 = h_2$ in $D_h$, which imply the 2-3 permutation symmetry,  
this matrix gives
$$
m^{ls}_\nu  \propto
\begin{pmatrix}
2u_1(h_1+2h_2)  & (u_1+u_2)(h_1-h_2) &  (u_1+u_2)(h_1-h_2)\\
...    &  2u_2(h_1+2h_2)     & 2u_2(h_1-h_2)\\
... & ...   &  2u_2(h_1+2h_2)
\end{pmatrix},  
$$
with  maximal 2-3  and zero 1-3 mixings. 
Additional assumption $h_2 = -h_1$ is then required 
to produce  the TBM mixing, since $U_l = {\mathbbm 1}$. 
Three parameters $h_1$, $u_1$ and $u_2$ allow to get arbitrary masses for the three light neutrinos.

\subsection{Extended hidden sector}
\label{chapt:exte}

Introduction of more than three fermionic singlets of SO(10)  
opens up new possibilities to obtain the observed mixing and masses too. And
in some cases it allows to simplify the conditions on couplings and VEVs we 
imposed in the previous sections.  On the other hand,  the  Dirac screening may not be 
straightforward any more.

Let us consider for definiteness effects of three additional SO(10) 
singlets,
so that the hidden fermion sector consists of 
$S\,\sim\,({\bf 3}, \vec{\bf 1})$ of $A_4$.
The light  neutrino  mass matrix is given by
\begin{equation} 
\label{more:singlets}
m_\nu = (m_D)_{3\times 3}\cdot {1\over (M_D)_{3\times n} \cdot
(m_S^{-1})_{n \times n} \cdot (M_D^T)_{n \times 3}} \cdot 
(m^T_D)_{3\times 3},
\end{equation}
where $n=6$.
Taking into account the $A_4$ symmetry assignment we can write
different  mass matrices involved as
\begin{equation}
m_D = M^{(D)}_{33},\, ~~~ M_D =  
\begin{pmatrix}
   M^{(P)}_{33}, M_{31}^{(P)}
\end{pmatrix},\, ~~~M_S =
\begin{pmatrix}
M^{(S)}_{33}    & M_{31}^{(S)}\\
M_{13}^{(S)}    & M^{(S)}_{11}
\end{pmatrix}.
\label{3matrices}
\end{equation}

Consider diagonalization of the inverse matrix $[m_\nu]^{-1}$.  
Recall that if $m_\nu$ is diagonalized by the unitary matrix $U$, 
then $[m_\nu]^{-1}$ is diagonalized 
by $U^*$. Consequently, for a real $U$ as in the case of $U_{TBM}$,  
$[m_\nu]^{-1}$ is diagonalized by the same matrix $U$. 
The inverted mass matrix can be written as 
\be
m_\nu^{-1} = m_D^{-1} M_N m_D^{-1}, 
\label{invmnu}
\ee
where $M_N$ is the effective mass matrix of RH neutrinos:
\be
M_N = M_D^T M_S^{-1} M_D.
\label{mnmatrix}
\ee
Using expressions  (\ref{3matrices}) we find 
in terms of $3 \times 3$ matrices:
\be
M_S^{-1} =
\begin{pmatrix}
\left[M_{33} - M_{31} M_{11}^{-1} M_{13} \right]^{-1}_S   &
\left[M_{13} - M_{11} M_{31}^{-1} M_{33} \right]^{-1}_S\\
\left[M_{31} - M_{33} M_{13}^{-1} M_{11} \right]^{-1}_S   &
\left[M_{11} - M_{13} M_{33}^{-1} M_{31} \right]^{-1}_S
\end{pmatrix}, 
\label{invms}
\ee
where the subscript $S$ means that all the matrices in the bracket 
belong to  the hidden sector. Inserting it into Eq.~(\ref{mnmatrix})  and then into
Eq.~(\ref{invmnu}) we obtain
\begin{eqnarray}
m_\nu^{-1} & = &  (M^{(D)}_{33})^{-1}
\left\{M_{33}^{(P)}\left[M_{33} - M_{31} M_{11}^{-1} M_{13} \right]^{-1}_S M_{33}^{(P)} \right.
\nonumber\\
& + &  M_{33}^{(P)}\left[M_{13} - M_{11} M_{31}^{-1} M_{33} \right]^{-1}_S M_{31}^{(P)\,T} +
M_{31}^{(P)}  \left[M_{31} - M_{33} M_{13}^{-1} M_{11} \right]^{-1}_S M_{33}^{(P)}
\nonumber\\
& + & \left. M_{31}^{(P)} \left[M_{11} - M_{13} M_{33}^{-1} M_{31} \right]^{-1}_S  M_{31}^{(P)\,T} \right\}
(M^{(D)}_{33})^{-1}. 
\label{invmnumnu}
\end{eqnarray}
Imposing various conditions on the matrices involved, 
one can get specific structures of $m_\nu^{-1}$ and lepton mixing.

Let us consider a special case of  $M_{13}^{(S)} \equiv (M_{31}^{(S)})^T = 0$, 
for which, 
 in fact, we can not use general formula (\ref{invmnumnu}) with the inverse of $M_{13}^{(S)}$. 
In this case   $M^{(S)}_{33}$
and $M^{(S)}_{11}$  contribute independently to the mass matrix of the RH neutrinos, 
{\it i.e.} the denominator of Eq.~\eqref{more:singlets}, through their own portal interactions. 
Plugging matrices (\ref{3matrices}) into  Eq.~(\ref{more:singlets}) we obtain 
\be
m_\nu = M^{(D)}_{33}~ { 1 \over M^{(P)}_{33} [M^{(S)}_{33}]^{-1} M^{(P)}_{33} 
+ M^{(P)}_{31} [M^{(S)}_{11}]^{-1} M^{(P)\,T}_{31}}~ M^{(D)}_{33},
\label{mnuzerod}
\ee
whose inverse  can be written as 
\be
[m_\nu]^{-1}   = 
[M^{(D)}_{33}]^{-1} M^{(P)}_{33} [M^{(S)}_{33}]^{-1} M^{(P)}_{33}[M^{(D)}_{33}]^{-1}
+
[M^{(D)}_{33}]^{-1} M^{(P)}_{31} [M^{(S)}_{11}]^{-1} M^{(P)\,T}_{31} [M^{(D)}_{33}]^{-1}. 
\label{mnuzerod2}
\ee   
In the first term the   
complete screening can be realized (as in the scheme of sect.~\ref{sect:tttt}) by
$$
[M^{(D)}_{33}]^{-1} M^{(P)}_{33} \propto {\mathbbm 1},
$$
if both matrices are generated by the same triplet $\xi$ only. 
Furthermore,  for the flavon triplet, its VEVs 
$\langle \xi \rangle = u(1,1,1)$, the singlet matrix in hidden sector 
$M^{(S)}_{33} = M_{33}^{special}$. Therefore 
the first term in Eq.~(\ref{mnuzerod2}) reduces to  $[M_{33}^{special}]^{-1}$. 
It is straightforward to see that the  inverse of $M_{33}^{special}$
has the same special form and therefore can be diagonalized by $U_{TBM}$. 

In the second term of Eq.~(\ref{mnuzerod2}) the partial screening  can be realized 
as in the scheme of Eq.\eqref{modelA:Mnu}: 
$[M^{(D)}_{33}] [M^{(P)}_{31}]^{-1} \propto  U^*_\omega$ or 
$[M^{(D)}_{33}]^{-1} M^{(P)}_{31} \propto  U_\omega$. 
Thus, expression (\ref{mnuzerod2}) reduces to 
\be
[m_\nu]^{-1} = [M_{33}^{special}]^{-1} + g~ U_\omega [M^{(S)}_{11}]^{-1} U_\omega,  
\label{more:zero31}
\ee
where $g$ is a coefficient,  
which takes care of the proportionality and normalization of matrices, 
thus  it is irrelevant for this discussion. 

Finally, the proportionality $M^{(S)}_{11} \propto {\mathbbm 1}$  can be imposed by certain symmetries that only apply to hidden sector interactions. 
In this case   
the second term  in Eq.~(\ref{more:zero31}) becomes 
$g U_\omega \, {\mathbbm 1} \, U_\omega$, which reproduces exactly 
the correction   $\Delta M$ as in  Eq.~\eqref{perturbms}. 
This correction removes ambiguity in diagonalization of the first 
term and fixes the TBM mixing. There are enough parameters left to reproduce 
the observed neutrino masses. 

\vspace{.15cm}

\subsection{On the CKM new physics} 
\label{chapt:CKM}

So far  we have discussed situations that no CKM mixing is generated 
and all the Dirac masses of SM 
fermions are the same: $m_u = m_d = m_l$, 
contradicting  observations. Additional physics 
(new fields, interactions) is therefore needed to explain 
properties  of the quark sector. This additional physics   
should not  destroy the constructions in the neutrino sector. 
Here we outline how this can be done, and refer to \cite{Ludl:2015tha} for more details.  
To be in agreement with observations, three aspects should be addressed. 

1).  Difference of the  up-  and  down- type fermions mass hierarchies.

There are various ways to  
obtain different mass hierarchies of the ``upper'' and ``down'' 
fermions (in the SM doublets) and generate the CKM mixing. 

(i)  A second 10-plet of Higgses, $10_H^d$, can be introduced to give 
masses of the down fermions,  whereas the first one, $10_H^u$, generates  
masses of the upper fermions:
up-type quarks and neutrinos. Eq.~\eqref{Lagr:general} then results in
\be
m_u = m_D = Y_{10}^u v_u, ~~~~ m_d = m_l = Y_{10}^d v_d. 
\label{Dirac:mass}
\ee
Therefore, different Yukawa coupling matrices $Y_{10}^u$ and $Y_{10}^d$    
produce different mass hierarchies and also CKM mixing. 

To keep the neutrino part unchanged we assume that 
in the symmetry basis the CKM mixing comes from the down fermions.
That is,  couplings of $10_H^d$ further violate $A_4$, explicitly or spontaneously. 

(ii) Another possibility is to keep the single $10_H$ but 
introduce the dimension-5 operator  
$$
{1\over \Lambda} (16_F  {16}_H {16}_H  16_F). 
$$ 
This operator also breaks the mass degeneracy 
between up- and down- type quarks \cite{Babu:1998wi}.\\ 

(iii) Vector-like families of fermions  $10_F$,  $\overline{10}_F$   
can  exist. Their mixing with usual $16_F$ break the degeneracy between up-type 
and down-type masses of quarks. The reason is that    
$10_F$ only contains down-type quarks and lepton doublets, but no up-type quarks. 
The interactions which provide this mixing  
can originate from renormalizable operators 
$10_{F} 16_{F} 16_H$. \\

2).  Difference of masses of down quarks and charged leptons. 
To breaks the mass degeneracy between quarks and leptons 
 one can  use non-renormalizable terms 
with  additional Higgs multiplets. One  option 
is to introduce $45$-plet Higgs scalars $45_H$  
and the  dimension-5 operator
\be
{1\over \Lambda} (16_F \cdot 10_H^d ~45_H  \cdot  16_F) 
\label{nonre:45H}
\ee
with $45_H$ and  $10^d_H$ only \cite{Chang:2004pb}. 
The product of $10^d_H$ and $45_H$ 
has the decomposition 
$ 10\times 45 = 10 + 120 + 320$, which contains $120_H$.  
If  $45_H$ acquires its VEV in the direction of ($B$-$L$), 
where $B$ ($L$) is the baryon  (lepton) number,  
then the effective  VEV of $120_H$ 
leads via interactions (\ref{nonre:45H}) to  difference of masses of the second fermion 
generation (e.g. $m_\mu$ and $m_s$).

If  more than one $45$-plet exist, 
the dimension-6 operators   
$$
{1\over \Lambda^2} (16_F  \cdot  45^a_H 10_H 45^b_H  \cdot  16_F)
$$ 
give  contributions to  the masses of the first generation 
of fermions \cite{Anderson:1993fe, Morisi:2007ft}. \\

3). Nearly the  same CKM-type mixing of down-type quarks and charged leptons despite the different mass hierarchies of them.   
In our framework   the CKM matrix is generated by $m_d$, and 
the CKM type correction to lepton mixing  
which lead to expression (\ref{ql:comp}) are generated by $m_l$. 
Since masses of the down-type quarks and charged leptons are different, 
the mass matrices $m_d$ and $m_l$ should be different, and therefore 
in general,  one expects that diagonalization matrices  $U_d$ 
and $U_l$ that diagonalize them are also different.   
This may invalidate Eq.~(\ref{ql:comp}), which implies  that 
both matrices should be  diagonalized by the same rotation 
$U_l \simeq U_d$~\cite{Ludl:2015tha}.

One solution to this problem was described before in sect.~\ref{sect:tttt}.  
The mass matrix can be the sum of a  diagonal matrix proportional to the unit matrix and an off-diagonal matrix: 
$$
m_0 {\mathbbm 1}  + M_{33}^{\xi}. 
$$ 
The mixing is determined by  $M_{33}^{\xi}$ only, whereas 
the eigenvalues are given  by the total matrix, which depends on  
$m_0$ and $M_{33}^{\xi}$.  
So, by varying $m_0$, we can change the eigenvalues without modifying the mixing.

\section{Conclusion} \label{sect:conclusion}

1. We have considered the lepton mixing and neutrino masses in the framework
of SO(10) GUTs with the hidden sector composed of fermionic and
bosonic  singlets of SO(10), and flavor symmetries realized in both visible and hidden sectors.
Two sectors communicate via the portal interactions provided by $\overline{16}_H$ of Higgses.

The framework allows to disentangle the CKM new physics
and the neutrino new physics. The neutrino new physics explains the smallness of
neutrino masses and generates the large lepton mixing.
The CKM new physics,  being  common for quarks and leptons due to GUT,
is responsible for the CKM mixing and  difference of
mass hierarchies of the  up-type and down-type quarks, as well as the  charged leptons.
The framework naturally realizes the relation between the quark and lepton mixings:
$U_{PMNS} \sim V_{CKM}^{\dagger} U_0$,  where the structure of matrix $U_0$ is
determined by the flavor symmetry.

In the lowest-order approximation with neutrino physics only: 
(i) quark and lepton masses are generated,
(ii) no quark mixing appears,
(iii)  large lepton mixing is produced,
whose form is dictated by the flavor symmetry,
(iv) small neutrino masses are generated via the double seesaw mechanism, 
or its generalizations are related to more than three fermionic singlets.

The CKM physics gives corrections (or perturbations)
to this lowest-order picture and
the  corrections are related to further violation of flavor
symmetry effects and higher-order operators. \\

2.  We have mainly focussed on the neutrino new physics and generation
of $U_0$ in this paper.
Using  $A_4\times Z_4$ as the  simplest example of flavor group, 
we have systematically explored the flavor structures
(masses, mixing) which can be obtained in this framework, 
depending on the field content and symmetry assignment.
The interplay of GUT SO(10) symmetry and flavor symmetries
is crucial in obtaining certain flavor structures.

In this framework three  different mass matrices are involved in generation of lepton mixing:
$m_D$ - the Dirac mass matrix of SM fermions (charged leptons and neutrinos),
$M_D$ - the portal mass matrix,
or $m_D M_D^{-1\,T}$ - the portal factor,  and the matrix of fermion singlets $M_S$.
This leads to new possibilities to explain the observed mixing. 
One of the key elements, which allows to disentangle
the quark masses and mixing from neutrino masses and mixing, 
is the Dirac screening (either complete or partial).

Generically, even without flavor symmetries, 
the framework explains no or small quark
mixing, large lepton mixing and small neutrino masses.
It leads to the relation between the quark and lepton mixings
of the form \eqref{ql:comp}.

$U_0$ in Eq.~\eqref{ql:comp} can be factorized as
$U_0 = U_1 U_2 ...$, where  $U_i$ are the matrices of maximal two-generation rotation or
matrices of special type,  such as the magic matrix $U_\omega$.
Then different mass matrices can be responsible for different rotations $U_i$.
In this connection we studied properties of matrices $m_D$, $M_D$,  $M_S$ 
generated by different SO(10) Higgs multiplets.
Depending on $A_4$ assignments for fermions they can be
of 5 types: $M_{33}^{diag}$, $M_{33}^{\xi}$, $M_{31}$, $M_{11}$  and 
$M_0 \cdot {\mathbbm 1} $.

For a complete set of flavons (triplet and singlets) with
arbitrary couplings and VEVs, any (symmetric)
structure of $M_{33}$ and $M_{11}$ (and consequently,
$m_D$ and $M_S$) and $M_{31}$ can be obtained.
As a consequence, the flavor structure of
light neutrino mass matrix is arbitrary.

The matrices acquire certain flavor structures 
if only specific flavons (e.g.  only triplet) contribute to their generation. If
the same flavons contribute to both $m_D$ and $M_D$,
the Dirac screening can be obtained.
This can be achieved by additional symmetry, e.g.   
$Z_4$. This symmetry forbids certain couplings of flavons,
ensuring that different flavons participate in
different types of interactions. It can then explains that some flavons appear in the hidden sector only.
Or on the contrary, a common symmetry assignment ensures that the same flavons contribute to
different mass matrices. 

It is also possible that certain symmetries exist in the hidden sector only, being broken
(spontaneously or explicitly) by low scale interactions in the visible
and portal interactions.

To have a complete determination of masses and mixing, in particular,
to obtain $U_0 \sim U_{TBM}$, or $U_{BM}$,
further restrictions are needed. We found 
the following generic  conditions,  of which all or some should be satisfied in specific 
schemes:

- equality (or proportionality) of the flavon singlet couplings $\vec{y}$ in the visible
sector ({\it i.e.} with $10_H$) and in portal
(with $\overline{16}_H$) interactions;

- equality of  flavon singlet VEVs;

- correlation between VEVs of flavon singlets and triplet;

- certain VEV alignment of triplet(s) $\xi$ 
including equality of three VEVs, or
non-zero VEV of only one component of the triplet;

- corrections to special matrices which break
degeneracy of mass eigenstates and thus 
fix the mixing.

These relations can be manifestations of further
unification, additional symmetries, or 
selection of certain minimum of the potential
from several (degenerate) possibilities.
In some cases  correlations of the bare
mass terms or flavonless contributions 
are also required. \\

3.  We have presented several schemes with minimal  
possible number of flavons and couplings
in the visible and portal sectors.
Interesting realizations are
when $U_{TBM}$, $U_{BM}$, or certain factors of these
matrices follow from different sectors:

- the TBM mixing from the hidden sector
$M_S$ (with complete screening);

- approximate TBM mixing from charged leptons;

- approximate TBM mixing from the portal and hidden sectors;

- the magic matrix from charged leptons and maximal 1-3 mixing from
the hidden sector, thus reproducing $U_{TBM}$;

- the magic matrix from the portal (partial screening) and
maximal 1-3 mixing from the hidden sector;

- the 2-3 maximal mixing from charged leptons and
the 1-2 maximal mixing from $M_S$, thus reproducing $U_{BM}$.

In many cases explanations of masses and mixing decouple 
from each other. This is because of the special 
(maximal, magic) form of the mixing.

We also commented on possible effects of the linear seesaw and
additional fermion singlets in the hidden sector.
They can provide required
corrections to the zero-order structures.

We outlined the CKM physics, indicating  additional 
elements which could be introduced to obtain the quark sector. We showed that they can be introduced
in a consistent way without destroying  the part  related to neutrino new physics. \\

4. This is an initial study  -- the first step 
toward the model building, which shows, nevertheless,  
interesting potential to explain fermion masses and mixing.
In complete models one needs to further introduce the scalar potential explicitly and
find its minima, to consider effects of high-order operators, and to 
take into account renormalization group effects, {\it etc}. 
Other flavor symmetry groups can be studied. 
All these issues are beyond the scope of the present paper.\\

5 . The key challenge is the possibility of testing this framework experimentally. 
In some cases (or with additional assumptions) one can obtain
certain predictions for the neutrino mass hierarchy,
the type of mass spectrum,
value of the Dirac CP phase and values of Majorana phases. 
Discovery of the proton decay would be a partial confirmation of the GUT 
framework. The leptogenesis can give  some probes of the framework too.

The hidden sector can provide stable particles to act as dark matter candidates. 
So, detection of dark matter particles may shed some light on the hidden sector, 
and consequently, the framework studied in this paper. 
Sterile neutrinos, if exist,  can also originate from such a hidden sector.

\appendix

\section{$A_4$ group}  
\label{a4group}

In the context of $A_4$ flavor symmetry, we use decomposition of 
product of two triplets $a$, $b$ as follows 
\bea
{\bf 1}&=&(ab)=a_1 b_1 +  a_2 b_2 + a_3 b_3,
\nonumber\\
{\bf 1' }&=&(ab)'=a_1 b_1 +  \omega^2 a_2 b_2 + \omega a_3 b_3,
\nonumber\\
{\bf 1''}&=&(ab)''=a_1 b_1 +  \omega a_2 b_2 + \omega^2 a_3 b_3,
\nonumber\\
{\bf 3 }&=&a \cdot b= h_1 \cdot (a_2 b_3,  a_3 b_1, a_1 b_2)+ h_2 
\cdot (a_3 b_2, a_1 b_3, a_2 b_1),
\eea
where $\omega = e^{2\pi i/3}$, and $h_1$ and $h_2$ are both free number parameters. 
We further assume that $h_1 = h_2$ is protected 
by some underlying mechanism, and in practice choose the value of $h_1$ to be $1$ 
because it can be absorbed into relevant Yukawa couplings.  Generally, this assumption does 
not affect the forms of $m_D$ and $M_S$, which are required 
to be symmetric anyway. But in the case that both $\Fi$ and $S_i$ 
are triplets under $A_4$,  this assumption leads to a symmetric $M_D$,  and thus
helps to preserve the screening condition which links $M_D$ to $m_D$. 

\section{Eigenvalues of $M_{33}^{\varphi}$}
\label{app:eigenvalues}
Let us denote  $\mu \equiv m_0 + y v$,  $\mu' \equiv  y' v'$ and  
$\mu''  \equiv y'' v''$.  
The moduli of the eigenstates of $M_{33}^{\varphi}$ (\ref{33diag}) equal 
\begin{eqnarray}
|M_1| & = &|\mu + \mu' + \mu''|,
\nonumber\\
|M_2| & = & [ |\mu|^2 + |\mu'|^2 + |\mu''|^2 + 2\text{Re}(\omega^2 \alpha)]^{1/2},
\nonumber\\ 
|M_3| & = & [ |\mu|^2 + |\mu'|^2 + |\mu''|^2 + 2\text{Re}(\omega  \alpha)]^{1/2}. 
\end{eqnarray}
and  $\alpha \equiv  \mu \mu'^* + \mu' \mu''^* + \mu'' \mu^*$.  
The arguments of $M_ 2$  and $M_3$  are given by 
$$
2\arctan[{ \text{Im}(2\mu - \mu' -\mu'') \pm \sqrt{3}\text{Re}(\mu'-\mu'')  
\over |2M_{2,3}| + \text{Re}(2\mu - \mu' -\mu'') \pm \sqrt{3}\text{Im}(\mu''-\mu') }],
$$ 
respectively.  If $\alpha$ is real, the moduli of two eigenstates are degenerate.

\section{Diagonalization of $M_{33}^{\xi}$}  
\label{diag:off}

Consider the diagonalization of the real matrix
\begin{equation}
M_{33}^{\xi} = 
\begin{pmatrix}
0  & u_3 &  u_2\\
u_3      & 0    & u_1\\
u_2 & u_1  &  0
\end{pmatrix}.
\end{equation}
Its eigenvalues  are given by 
\begin{eqnarray}
m_1 &=& {2\over 3 } \sqrt{3u_1^2 + 3u_2^2 + 3 u_3^2}  \cdot \cos{\Delta - 2\pi \over 3},\notag\\
m_2 &=& {2\over 3 } \sqrt{3u_1^2 + 3u_2^2 + 3 u_3^2} \cdot \cos{\Delta + 2\pi  \over 3}, \notag\\
m_3 &=& {2\over 3 } \sqrt{3u_1^2 + 3u_2^2 + 3 u_3^2}\cdot \cos{\Delta \over 3},
\end{eqnarray}
where
$$
\cos \Delta \equiv  {27 u_1 u_2 u_3 \over (3u_1^2 + 3u_2^2 + 3 u_3^2)^{3/2}}. 
$$
If the elements are nearly degenerate as $u_i = u + \epsilon_i$, where $\epsilon_i \ll u$, we have
\begin{eqnarray}
m_1 &\simeq & -u - {1\over 3 } \left(\epsilon_1 + \epsilon_2 +\epsilon_2  -
2 \sqrt{\epsilon_1^2 +  \epsilon_2^2 +\epsilon_3^2 - \epsilon_1\epsilon_2 -
\epsilon_2\epsilon_3 - \epsilon_3\epsilon_1} \right),\notag\\
m_2 &\simeq  & -u - {1\over 3 }  \left(\epsilon_1 + \epsilon_2 +\epsilon_2 +
2 \sqrt{\epsilon_1^2 +  \epsilon_2^2 +\epsilon_3^2 - \epsilon_1\epsilon_2 - \epsilon_2\epsilon_3 -
\epsilon_3\epsilon_1} \right) , \notag\\
m_3 &\simeq & 2 u+ {2\over 3 } \left(\epsilon_1 + \epsilon_2 +\epsilon_2\right).
\end{eqnarray}

Under the condition that $|u_3| \le u_2 \le u_1$, it is possible to give
the three mixing angles\,\cite{Kronenburg:2013na}:
\begin{eqnarray}
\cos^2 \theta_{13} &=& {2 m_3^2 - u_3^2 - u_2^2 \over (m_3 - m_1)(m_3 - m_2 ) }
,\notag\\
\sin^2 \theta_{12} & = & {(m_3-m_1) \cos^2\theta_{13} - m_3 \over (m_2 - m_1)\cos^2\theta_{13} }
,\notag\\
\cos 2\theta_{23} &=&  {(m_1-m_2)\sin\theta_{13}\sin2\theta_{12} 
\over 2 u_1} .
\end{eqnarray}
Note that the mass matrix always  can be redefined to satisfy this condition.

\section{the TBM and BM mixings} 
\label{App:mixings}

The TBM mixing 
\be
U_{TBM} = 
\begin{pmatrix}
\sqrt{ 2\over 3}   &  {1\over \sqrt{3}}  &  0\\
-{1\over \sqrt{6}}       & {1\over \sqrt{3}}    & -{1\over \sqrt{2}} \\
-{1\over \sqrt{6}}       & {1\over \sqrt{3}}  & {1\over \sqrt{2}} 
\end{pmatrix},
\ee
can be generated by the physical neutrino mass matrix in the flavor basis 
\begin{equation} 
m_\nu  =
\begin{pmatrix}
x  + z    & y  &  y\\
y      &z   & x +y \\
y      &x + y  &z
\end{pmatrix},  
\end{equation}
whose form is referred to as $M_{TBM}$ in the main text. Here $x$,\,$y$,\,$z$ are  free parameters.  
How to obtain TBM in other bases have been investigated in sect.~\ref{TBM:Factorizing} above.

The BM mixing 
\be
U_{BM} = 
\begin{pmatrix}
{1\over \sqrt{2}}   & -{1\over \sqrt{2}}   &  0\\
{1\over 2}     &{1\over 2}  & -{1\over \sqrt{2}}\\
{1\over 2}      &{1\over 2}  & {1\over \sqrt{2}}
\end{pmatrix},
\ee
corresponds  to the neutrino mass matrix  in the flavor basis
\begin{equation} 
m_\nu  =
\begin{pmatrix}
x     & y  &  y\\
y      &z   & x-z\\
y      &x-z  &z
\end{pmatrix}.  
\end{equation}

\end{document}